\em

\documentclass[a4paper,10pt]{article}   
\usepackage{lscape,a4wide,amssymb,amsmath,cite,epsfig, psfrag}  
\newcommand{\be}{\begin{equation}}   
\newcommand{\ee}{\end{equation}}   
\newcommand{\bea}{\begin{eqnarray}}   
\newcommand{\eea}{\end{eqnarray}}   
\newcommand{\bean}{\begin{eqnarray*}}   
\newcommand{\eean}{\end{eqnarray*}}

\newcommand{\gapproxeq}{\lower   
.7ex\hbox{$\;\stackrel{\textstyle >}{\sim}\;$}}   
\newcommand{\lapproxeq}{\lower   
.7ex\hbox{$\;\stackrel{\textstyle <}{\sim}\;$}}

\newcommand{\la}{\langle}   
\newcommand{\ra}{\rangle}   
\newcommand{\lra}{\leftrightarrow}   
\newcommand{\bc}{\begin{center}}   
\newcommand{\ec}{\end{center}}   
\newcommand{\btab}{\begin{tabular}}   
\newcommand{\etab}{\end{tabular}}

\def\qq{$ q\bar q $}   
\def\ss{$ s\bar s $}   
\def\nn{$n\bar{n}$}

\def\cc{$c\bar c$}     
\begin{document}   
   
\begin{titlepage}   
   
  \baselineskip=18pt \vskip 0.9in
\begin{center}   
  {\bf \Large Hybrid meson production by electromagnetic and weak
    interactions
    in a Flux-Tube simulation of Lattice QCD}\\
  \vspace*{0.3in} {\large F.E. Close}\footnote{\tt{e-mail:
      f.close@physics.ox.ac.uk}}
  {\large and J.J.Dudek}\footnote{\tt{e-mail: dudek@thphys.ox.ac.uk}}\\
  \vspace{.1in}
  {\it Department of Physics - Theoretical Physics, University of Oxford,\\
    1 Keble Rd., Oxford OX1 3NP, UK} \\
  \vspace{0.1in}
\end{center}   
   
\begin{abstract}   
  We calculate rates for hybrid meson production by electromagnetic
  and weak interactions in the flux-tube model.  Applications include
  photo and electroproduction at Jefferson Laboratory and HERA, and
  the production of light strange and charmed hybrids in the weak
  decays of heavy flavours.  Photoproduction of some light hybrids is
  predicted to be prominent in charge exchange reactions, $\gamma p
  \to n \cal{H}$ and accessible in $\gamma p \to p \cal{H}$.
  Production of light or charmed hybrids in $B$ and $D$ decays may be
  feasible with high statistics. Photoproduction of the axial hybrid
  meson is predicted to be large courtesy of $\pi$ exchange, and its
  strange counterpart is predicted in $B \to \psi K_H(1^+)$ with
  $b.r. \sim 10^{-4}$. Production rates for 
  exotic hybrid candidates $1^{-+};(0,2)^{+-}$ are given special
  attention. Selection rules that can help to distinguish between
  hybrid and conventional states with the same $J^{PC}$ are noted.
   
\end{abstract}   
\end{titlepage}

\subsection*{Introduction}

An outstanding problem in the Standard Model is how the non-Abelian,
gluon, degrees of freedom behave in the limit of strong QCD. Lattice
QCD predicts a spectroscopy of glueballs \cite{glue} and hybrid mesons
\cite{hybrids}, but there are no unambiguous signals against which
these predictions can be tested.
   
A major stumbling block in the case of hybrids is that while
predictions for their masses\cite{hybrids,IP}, hadronic
widths\cite{IK,CP95} and decay channels\cite{IK,CP95,michael2000} are
rather well agreed upon, the literature contains no general discussion
of their production rates in electromagnetic or weak interactions.
Meanwhile a significant plank in the proposed upgrade of Jefferson
Laboratory is its assumed ability to expose the predicted hybrid
mesons in photo and electroproduction.
   
Clearly a calculation of hybrid photoproduction in a model based on
lattice QCD is urgently called for. Theory\cite{IP,conf} has provided
compelling arguments from QCD that confinement occurs via the
formation of a flux tube. In the simplest situation of a long tube
with fixed $Q,\bar{Q}$ sources on its ends, a flux-tube has a simple
vibrational spectrum corresponding to the excitation of transverse
phonons in its string-like structure. There is a question whether the
flux-tube is fully formed on the ${\cal O}(1)$ fm scale typical of
hadrons (ref. \cite{kuti} suggests that at ${\cal O}(1)$ fm the state
of confined gluon fields is between that typical of bag models and
those of a fully formed flux tube; by contrast \cite{thomas} finds
that the flux-tube forms at distances below 1 fm). The model of
ref.\cite{IP} assumes that the fully formed tube drives the
phenomenology and that the essential features of this gluonic spectrum
are retained in the spectrum of real mesons with their flux-tube
excited - the hybrid mesons.

The flux tube model has not been derived from QCD, but it has many
features that appear to be shared by lattice QCD. As such it is
currently the best that we have that builds on features emerging from
the strong coupling limit of QCD. Hopefully its successes or failings
can enable deeper insights to emerge as to the dynamics of QCD in the
strong interaction limit. It leads to the effective linear potential
of the conventional meson spectroscopy and has an identical N=1
spectrum of hybrid states as the lattice - a feature that is
especially relevant to the present calculations. Its pedigree includes
predictions that have inspired studies on the lattice and even
anticipated their results (such as the $J^{PC}$ pattern of the
lightest hybrids \cite{IP}, and that the decays of those hybrids to
ground state mesons are suppressed relative to those to excited states
\cite{IP} \cite{CP95}, which has recently been confirmed in the
lattice framework \cite{michael}).  We would hope that insights from
the present work in their turn might inspire future lattice study,
which could thereby better establish their connection with reality.

The model is widely used in the experimental community for whom the
reasons underlying the lattice results are often obscure. As such it
is playing a prominent role in the the Jefferson Laboratory upgrade
proposal, where the flux-tube idea has been used to underpin much of
their planning. To turn this into something of practical use will
require predictions for the electromagnetic transition amplitudes to
hybrid mesons. However, while implications for spectroscopy and hadronic decays in
such a model have been extensively explored, previous estimates of
electromagnetic couplings in this model (refs.\cite{CP97,pageanas})
are at best only upper limits, in that they were based upon vector
meson dominance of hadronic decays of hybrids into modes including
$\rho$ and further assumed that the predicted suppression of decay into
$\pi \rho$ is suspended in $\pi$ exchange.  We are unaware of any
direct calculation of photo- or electro-production of hybrids in this
model.  This is the issue that we address in this paper.
   
From among the results of our extensive survey we note the following.
 
(1) The electric dipole transitions of the hybrid axial meson to
$\pi^{\pm} \gamma$ and of the exotic $(0,2)^{+-}$ to $\rho \gamma$
give radiative widths that can exceed 1 MeV. This implies significant
photoproduction rates in charge exchange reactions $\gamma p \to H^+
n$.  The exotic $2^{+-}$ may also be produced diffractively in $\gamma
p \to H p$.
   
(2) The production of an axial strange hybrid in $B \to \psi K_H$ is
predicted to have $b.r. \sim O(10^{-4})$ so long as its mass $\leq 2.1
\mathrm{GeV}$. There is a tantalising unexplained enhancement in the
data that may be compatible with this\cite{psidata} and merits further
investigation.
   
(3) If there is a light exotic hybrid\cite{e852,ves,vesb1,E852a} with
$J^{PC} = 1^{-+}$, the I=1 and I=0 states could be produced in $D \to
\pi H$ and $D_s \to \pi H$ with branching ratios $\sim 10^{-7}$.
   
(4) Measurement of the axial to vector amplitude ratio in $B \to
D_{(H)} X$ may enable the hybrid content of charmed mesons to be
determined.
   
\section{The model \label{model}}

The flux-tube is a relativistic object with an infinite number of 
degrees of freedom.  A standard approximation\cite{IP,IK,CP95,isgur99} 
has been to fix the longitudinal separation of the $Q\bar{Q} \equiv r$ 
and to solve the flux-tube dynamics in the limit of a thin 
relativistic string with purely transverse degrees of freedom. The 
resulting energies $E(r)$ are then used as adiabatic effective 
potentials on which the meson spectroscopies are built. Ref.\cite{bcs} 
studied the effect of relaxing these strict approximations and found 
that the spectrum of the conventional and lowest hybrids is robust. We 
shall assume the same is true in this first calculation of 
electromagnetic excitation of hybrid mesons. 
   
In refs.\cite{IP,isgur99,bcs} the flux-tube was discretised into $N+1$ 
cells, (modern lattice computations typically have $N \sim 10$) and 
then $N \to \infty$. Up to $N$ modes may be excited.  We shall focus 
on the first excited state, with excitation energy $\omega = 
\frac{\pi}{r}$ where $r$ is the length of the flux tube.  If the 
length of a cell is $l$ then $r = (N+1)l$. 
   
The state of the flux tube can be written in terms of a complete set 
of transverse eigenstates 
\[   
| \vec{y} \rangle = | \vec{y}_1 \ldots \vec{y}_n \ldots \vec{y}_N 
\rangle 
\]   
and the Fourier mode for the first excited state\footnote{Higher modes 
  (up to $p=N$) exist but we are not interested in them here as we 
  wish to specialise to the lightest hybrid mesons. Incorporating such 
  modes is straightforward in the formalism we describe.} is \be 
\vec{a} = \sqrt{\frac{2}{N+1}} \sum_{n=0}^{N}\vec{y}_n \sin \frac{\pi 
  n}{N+1} 
\label{asum}  
\ee or \be \vec{y}_n = \sqrt{\frac{2}{N+1}} \; \vec{a} \sin \frac{\pi 
  n}{N+1} 
\label{ysum}  
\ee The oscillations are in the two dimensional space transverse to 
the nominal $Q\bar{Q}$ axis. Thus there are two Fourier modes $\vec{a} 
\equiv (a_1,a_2)$ where $1,2$ refer to the two (body-fixed) orthogonal 
coordinate directions, $\hat{e}_1, \hat{e}_2$. 
   
In the small oscillation approximation the system becomes harmonic in 
$\vec{y}$ ($\vec{a}$). 
Then if $b$ is the string tension ($b \sim 1$ GeV/fm), the 
eigenfunctions for the ground and first excited states (labeled $0,1$ 
respectively) are in the Fourier-mode space 
   
\begin{eqnarray}  
\chi_0(a_{1,2}) &=& \left(\frac{b}{N+1}\right)^{1/4} \exp \left[-  
  \frac{b \pi}{2(N+1)} a_{1,2}^2 \right] \label{chi1}\\  
\chi_1 (a_{1,2}) &=& \sqrt{\frac{2 b \pi}{N+1}} \; a_{1,2}  
\chi_0(a_{1,2}) \label{chi2}  
\end{eqnarray}

To reduce the number of indices write $\vec{a} \equiv (a_1,a_2)$ in 
the body-fixed basis and understand any $\vec{a}$ to refer always to 
these components. To proceed to the continuum, write (see \cite{IP,isgur99}) 
\[   
\frac{ b \pi}{N+1} \equiv \beta_1^2 
\]  
  
The Gaussian wavefunctions eqns.(\ref{chi1}, \ref{chi2}) for the flux-tube 
ground-state and first excited mode become 
   
\begin{eqnarray}  
\chi_0(\vec{a}) &=& \left(\beta_1^2 /\pi \right)^{1/4} \exp \left[ - \beta_1^2 \vec{a}^2 /2 \right] \label{newchi1} \\  
\chi_1 (\vec{a}) &=& \sqrt 2 \beta_1 \vec{a} \chi_0(\vec{a}) \label{newchi2}  
\end{eqnarray}   
   
   
The wavefunctions for mesons must include the state of the flux-tube. 
So for conventional mesons, where the flux-tube is in its ground 
state, write 
$$ 
{\cal C} = \psi^{(0)}_{nlm}(\vec{r})\chi_0(a_1) \chi_0(a_2), 
$$ 
where $n,l,m$ are the usual two-body quantum numbers and the subscript zero indicates the ground state.
If either of the transverse modes is excited, one has a state that 
we refer to as a hybrid meson.  The particular combinations 
$\frac{1}{\sqrt{2}} [a_1 \pm i a_2] \equiv \frac{1}{\sqrt{2}} a^\pm$ 
give normalised circularly polarised phonon modes for the flux-tube, 
which have angular momentum $\pm 1$ about the longitudinal axis.  The 
corresponding wavefunction for such a hybrid may be summarised by 
$$ 
{\cal H} = \psi^{(\pm)}_{nlm}(\vec{r})\frac{1}{\sqrt{2}} 
\left[\chi_1(a_1) \chi_0(a_2) \pm i\chi_0(a_1) \chi_1(a_2)\right] 
$$ 
or 
$$ 
{\cal H} = \psi^{(\pm)}_{nlm}(\vec{r}) \; \beta_1 \left(a_1 \pm i 
  a_2\right) \chi_0(a_1) \chi_0(a_2). 
$$ 
   
The challenge is : how can electromagnetic or weak currents, which 
couple to quarks, break the orthogonality of $\la \vec{a} \ra$ needed 
to give a transition between conventional and (first) excited 
flux-tube (hybrid) state? The answer is implicit in the observation of 
Isgur\cite{isgur99} that the flux tube is a dynamic entity, with a 
zero-point motion which can affect observables that at first sight are 
driven by the $Q$ or $\bar{Q}$. 
  
The physical picture becomes transparent if one simulates the tube as 
a series of beads with mass $m$ on a massless string, and simplifies to the cases of 
$N=1,2$\cite{cdhugs}.  We will be interested here in excitation of the 
first hybrid mode, whose essential spatial structure can be simulated 
by a single bead of mass $m \equiv br$.

\begin{figure} 
\begin{center} 
  \psfragscanon \psfrag{rd}{$\frac{m_Q + \frac{b r}{2}}{m_T} \vec{r}$} 
  \psfrag{rQ}{$\frac{m_d + \frac{b r}{2}}{m_T} \vec{r}$} 
  \psfrag{a}{$\sqrt{\frac{2}{b\pi}}\beta_1 \vec{a}$} 
  \psfrag{rtransQ}{$\frac{\beta_1 r}{m_Q} \sqrt{\frac{2 b}{\pi^3}} \vec{a}$} 
  \psfrag{rtransd}{$\frac{\beta_1 r}{m_d} \sqrt{\frac{2 b}{\pi^3}} \vec{a}$}
\includegraphics[width=3in]{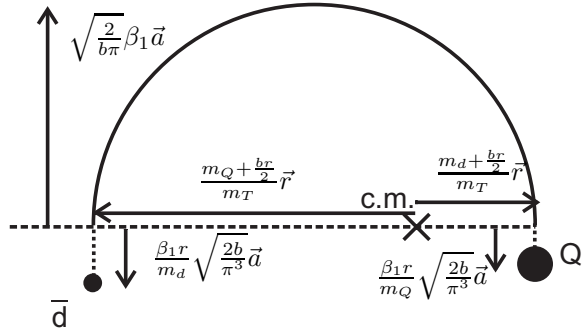}
\caption{$p=1$ mode hybrid structure. $m_T=m_q+m_d+br$. \label{fluxtube}}  
\end{center} 
\end{figure}

In this first excited mode the centre of mass of the 
$Q\bar{Q}$-bead system is displaced from the interquark axis by a 
transverse distance that scales as $ \sim m/m_Q$.  If the transverse 
displacement of the bead is $\vec{y}$, 
and the $Q$ and $\bar{Q}$ have masses $m_Q$, then relative to the 
centre of mass, the position vector of the quark (antiquark) has 
components in the longitudinal $\vec{r}$ and transverse $\vec{y}$ 
directions 
\be 
\vec{r}_{Q(\bar{Q})} = \left[\pm\frac{1}{2}\vec{r} \; ; 
  \left(\frac{br}{2 m_Q}\right) \vec{y}\right] 
\label{rvec1}  
\ee 
The dependence of $\vec{r}_{Q(\bar Q)}$ on $\vec{y}$ enables a 
quark-current interaction to excite transitions in the $\vec{y}$ 
oscillator, leading to excitation of the flux-tube.  The presence of 
$\pm$ in the $\vec{r}$ coordinate, but only $+$ in the case of 
$\vec{y}(\vec{a})$, is a feature of the first excited mode.  This is 
illustrated for the full flux-tube in fig.(\ref{fluxtube}). The longitudinal axis passes 
through the c.m. of the system. If the tube's effective c.m. (the 
``bead") is displaced transverse to this in one direction, then the $Q 
$ and $\bar{Q}$ respond collectively to the displacement of the flux 
tube, and are both on the opposite side of the longitudinal axis. 
Hence the same sign appears in the $\vec{y}(\vec{a})$ coordinate, but 
opposite signs in the longitudinal $\pm \vec{r}$. This sign will have 
significant consequences when we discuss $E1$ transitions. 
   
This is the essential physics behind the excitation of hybrid modes by 
current interactions with the quark or antiquark. 

Extending to $N$ beads\cite{isgur99} leads to more mathematical detail 
and enables excitations of up to $N$ modes, but the underlying 
principles are the same.  The position vector becomes \be 
\vec{r}_{Q(\bar{Q})} = \vec{R} \pm \frac{1}{2} \vec{r} + \vec{a} 
\frac{br}{\pi m_Q} \sqrt{\frac{2}{N+1}} 
\label{rvecn}  
\ee or in the continuum limit 
  
\be \vec{r}_{Q(\bar Q)} = \vec{R} \pm \frac{\vec{r}}{2} + \vec{a} 
\frac{r}{m_Q} \beta_1\sqrt{\frac{2b}{\pi^3}} 
\label{risgur}  
\ee 
  
The essential physics is already contained in the above examples. For 
modes with $p=\mathrm{even}$, the moment of the tube deformation relative to the 
interquark axis tends to cancel, leading to a null displacement of the 
quarks. For $p=\mathrm{odd}$, the tube has a net transverse moment, leading to a 
compensating transverse displacement of the quark(s).  This is encoded 
in the factor $(-1)^p$ for the quark displacement in the $p$-th mode 
in Isgur's formulation (eqn.(16) of \cite{isgur99}). 
 
Isgur\cite{isgur99} noted that the dynamical degree of freedom 
implicit in the $\vec{y}$ (or $\vec{a}$) gives a hitherto overlooked (
but welcome) contribution to the charge radius in elastic form 
factors. This comes about because the Coulomb interaction with the 
quark $F_{\mathrm{el}}(\vec{q}) = \langle g.s.| e^{i\vec{q}\cdot 
  \vec{r}_Q}|g.s. \rangle$ receives contributions from both the 
$\vec{r}$ and the $\vec{a}$ degrees of freedom at ${\cal O}(q^2 r^2)$ 
and ${\cal O}(q^2 a^2)$. 
  
In the continuum limit, $N \to \infty$ this becomes (compare eqns.(29) {\em et seq} in Isgur) 
\be 
F_{\mathrm{el}}(\vec{q}) = 1 - \frac{|\vec{q}|^2\la 
  r^2\ra}{24}(1+ \frac{8b}{ m_Q^2\pi^3}\sum_p \frac{1}{p^3}) 
\label{r2}  
\ee 
where the sum is over the $p = 1 \cdots \infty$ ``phonon" modes 
(elsewhere in most of this paper we consider only the $p=1$ mode).  It 
was shown\cite{isgur99} that these ``transverse excursions" give large 
$\sim 51\%$ corrections in light quark systems where $m_Q = m_d$, and 
$\sim 13\%$ corrections in heavy-light $Q\bar{q}$ systems. 
Furthermore the $\sum_1^\infty (1/p^3)$ is $\sim 80 \%$ saturated by 
its $p=1$ term.  Together, these suggested that the transition 
amplitudes to the lowest hybrids ($p = 1$ phonon modes) could be 
substantial. 
  
This is our point of departure. Expanding the incoming plane wave to 
leading order in the momentum transfer, 
\be \exp \left[-i\vec{q}\cdot\left(\frac{1}{2} \vec{r} + 
    \vec{a}\frac{r}{m_Q} \beta_1 \sqrt{\frac{2b}{\pi^3}} 
  \right)\right] \rightarrow \left(1 - i \frac{1}{2} \vec{q} \cdot 
  \vec{r}\right)\left(1 - i \vec{q}\cdot \vec{a} \frac{r}{m_Q} \beta_1 
  \sqrt{\frac{2b}{\pi^3}}\right) 
\label{pwave1}  
\ee the linear terms in $\vec{r}$ and $\vec{a}$ break the 
orthogonality of initial and final wavefunctions and cause 
transitions: $E1$ excitation to $L=1$ conventional states in the case 
of $\vec{r}$ and excitation of the first flux-tube mode (hybrid) in 
the case of $\vec{a}$. 
By combining the above with the tensor decomposition of the 
current-quark interaction, we may calculate electromagnetic and weak 
excitation amplitudes to hybrids, and compare with those for 
conventional mesons in various multipoles. 
   
First we take the continuum limit, define wavefunctions for both 
hybrid and conventional states including the flux-tube, illustrate the 
familiar $E1$ electromagnetic transition and then calculate its 
analogue for hybrid excitation with specific reference to the angular 
integrations. The electroweak transitions of heavy flavours will then 
be described, this requires knowledge of the decomposition of 
$V_{\mu}$ and $A_{\mu}$ for heavy light systems. 

When $m_Q \neq m_{\bar{Q}}$ Isgur's 
decomposition of the position vectors\cite{isgur99} was 
 \begin{equation}  
 \vec{r}_{Q(d)} = \vec{R} \pm \vec{r}\frac{\mu}{m_{Q(d)}} + \vec{a} 
 \frac{\beta_1 r}{m_{Q(d)}} \sqrt{\frac{2b}{\pi^3}}. \label{isgr} 
\end{equation}
where $\mu \equiv m_Q m_d/(m_Q + m_d)$.

  

  
For nomenclature we shall adopt the PDG\cite{PDG} notation, where the subscript 
$J$ denotes the total angular momentum of the meson, and we append the 
subscript $H$ to denote hybrid (thus for light flavours we have the notation 
in Table.(\ref{names}), with obvious generalisation for flavoured states).  
This is done for two reasons 
 
(i) to enable the trivial distinction between hybrid and conventional states 
to be immediately apparent and reduce confusion in the text; 
 
(ii) as a reminder that for a conventional and hybrid meson with the same overall 
$J^{PC}$, their internal \qq~ spin states are inverted.

\begin{table}[h] 
  \begin{center} 
    \begin{tabular}{cc|cc} 
      $J^{PC}_H$ & $S_{qq}$ & $I=1$ &$I=0$\\ 
      \hline 
      $1^{++}$ & $0$ & $a_{1H}$ & $f_{1H}$ \\ 
      $1^{--}$ & $0$ & $\rho_H$ & $\omega_H $\\ 
      $(0,1,2)^{+-} $& $1$ &$ b_{JH}$ & $h_{JH}$ \\ 
      $(0,1,2)^{-+}$ &$1$& $\pi_{JH}$ & $\eta_{JH}$ 
    \end{tabular} 
    \caption{Naming convention for light quark hybrid mesons 
    \label{names}} 
  \end{center} 
\end{table}

 This spin inversion is also illustrated in Table.(\ref{names}) and has potentially important 
 implications in helping to isolate hybrid contributions to the wavefunction of conventional states. 
 For example, whereas a meson with $J^{PC} = 1^{-+}; (0,2)^{+-}$ has exotic correlations of 
 $J^{PC}$ inaccessible to conventional mesons, the other states could a priori be  
 conventional or hybrid. Note however that the $1^{\pm \pm}_H$ have $S_{qq} = 0$ whereas 
 their conventional counterparts $1^{\pm \pm}$ have $S_{qq} = 1$. Conversely, the 
 $(0,2)^{-+}_H$ and $1^{+-}_H$ all have $S_{qq} = 1$ whereas their conventional counterparts 
 all have $S_{qq}=0$. Hence there is a complete spin inversion between conventional 
 states and their hybrid counterparts. 
  
 This spin inversion enables a dynamical distinction to ensue between these two 
 types of state, which is manifested in certain selection rules. This has already 
 been noted in hadronic decays\cite{CP95} and will have consequences in current transitions 
 too. Furthermore, we will find that the hierarchy of spin 
 operators leading to certain heavy flavour decays is inverted between conventional and hybrids, such  
 that observable consequence can ensue in principle.

\section{ E1 Transitions}   
   
\subsection{Conventional $Q\bar{Q}$ Transitions}   
   
To establish notation and make subsequent analysis of  
hybrid excitation more transparent, we first illustrate  
conventional E1 transition in the non-relativistic limit.   
Consider the  
$E1$ transition operator  
\[   
{\cal O}_{E1} = -i|\vec{q}| \sum_{q=Q,d} \vec{\epsilon}_+ \! \cdot  
(e_q\vec{r}_q)  
\]   
whose essential structure when acting on meson  
$Q\bar{d}$ is  
 \[   
 {\cal O}_{E1} = -i|\vec{q}|\left\{  
   \left(\frac{e_Q}{m_Q}-\frac{e_d}{m_d} \right)\mu \;  
   \vec{\epsilon}_+ \!\cdot \vec{r} + \left(  
     \frac{e_Q}{m_Q}+\frac{e_d}{m_d}\right) \sqrt{\frac{2b}{\pi^3}}  
   \beta_1 r \; \vec{\epsilon}_+ \! \cdot \vec{a}\right\}  
\]   
   If $l,m$ denote the orbital angular  
momentum of the $Q \bar{d}$ system, and its $z$-projection (on fixed  
space axes) respectively, then the structure of the matrix element  
$\cal{M}$ becomes  
\[  
{\cal M} \equiv \langle l',m' | {\cal O}_{E1} | l,m \rangle = -i  
|\vec{q}| \left( \frac{e_Q}{m_Q} - \frac{e_d}{m_d} \right) \mu   
\int \! d^3\vec{r} \int \! d^2\vec{a} \;  
\psi^{(0)*}_{n'l'm'}(\vec{r})\chi^*_0(a_1) \chi^*_0(a_2) \;  
\vec{\epsilon}_+ \! \cdot \vec{r} \;  
\psi^{(0)}_{nlm}(\vec{r})\chi_0(a_1) \chi_0(a_2)  
\]   
The integration over $d^2\vec{a} \to 1$ and the standard integral over  
$d^3\vec{r}$ gives a transition from $l=0$ to $l'=1$ caused by the presence  
of $\vec{r}$.

Separating into radial and angular parts, $\psi_{nlm}(\vec{r}) \equiv  
R_n(r) Y_l^m(\Omega)$ and noting that  
 \[   
 \vec{\epsilon}_+ \! \cdot \vec{r} \equiv \sqrt{\frac{4\pi}{3}}\; r  
 Y_1^{+1}(\Omega).  
 \]  
 $\cal{M}$ becomes  
\[   
-i |\vec{q}| \mu \left( \frac{e_Q}{m_Q} - \frac{e_d}{m_d}  
\right)\sqrt{\frac{4\pi}{3}}\; {_f\langle r \rangle_i} \int \!  
d\Omega\; Y_{l'}^{m'*} Y_{1}^{+1} Y_{l}^{m}  
\]  
where $_f \langle r \rangle _i \equiv \int \! r^2 dr\, R_{n'}^*(r)\,  
r\, R_n(r)$.

This general formula becomes more transparent when applied to the case  
$l=0,l'=1$ for which  
\[   
{\cal M} = -i|\vec{q}| \mu \left( \frac{e_Q}{m_Q} - \frac{e_d}{m_d} \right)  
{_f \langle r \rangle _i} \frac{1}{\sqrt{3}} \delta_{m',+1}  
\]   
We are interested in the  
specific case:  
 \begin{equation}  
  \label{Mpib}   
 {\cal M} (\gamma \pi \lra b_1) = i\left( \frac{e_1 - e_2}{m_n}\right)  
 {_b\la r \ra_\pi} |\vec{q}_b| \frac{m_n}{2 \sqrt{3}} \delta_{m',+1}.  
  \end{equation}  
  
In general we can write the radiative width as  
\begin{equation}   
  \label{width}   
  \Gamma (A \to B \gamma) = 4 \frac{E_B}{m_A} |\vec{q}| \frac{1}{2 J_A + 1}    
  \sum_{m_J^A} {| {\cal{M}}(m_J^A, m_J^B = m_J^A + 1) |}^2   
\end{equation}   
where the sum is over all possible helicities of the  
initial meson, and the matrix element is understood to be for a positive helicity photon.

This corresponds to the familiar E1 transition formalism of atomic and  
nuclear physics as traditionally applied to $Q\bar{Q}$ systems.  
   
 Notice that $\left( \frac{e_Q}{m_Q} - \frac{e_d}{m_d} \right) $  
ensures charge-conjugation conservation;  
for charge-neutral systems the $Q\bar{Q}$ charges cancel but they are  
vectorially on opposite sides of the c.m.(``longitudinal" electric  
dipole moment).  Hence a non-vanishing E1 amplitude occurs between  
neutral systems (e.g. $\chi \to \gamma \psi$).  
   
\subsection{Transitions to Hybrids}

Transitions to hybrids in the first excited state of the flux-tube 
arise from the $\vec{a}$ component of the $Q,\bar{Q}$ position 
operators. 
   
We consider the general matrix element for transitions between a 
conventional meson and a first excited mode hybrid with tube 
oscillation polarisation $\pm 1$; 
\[   
{\cal M} \equiv \langle \mathrm{hyb}; \pm, m'| {\cal O} 
|\mathrm{conv}; l,m \rangle = \int \!d^3\vec{r}\! \int\! d^2\vec{a} 
\;{\cal H^*OC} 
\]   
where (i) ${\cal O}$ - the essential spatial structure of the 
transition operator, which in this example is 
\begin{equation} 
{\cal O}_{E1} = -i |\vec{q}| \left( \frac{e_Q}{m_Q} + \frac{e_d}{m_d}  
\right) \sqrt{\frac{2b}{\pi^3}} \; \beta_1 r\; \vec{\epsilon}_+  
\!\cdot \vec{a} 
\label{opE1}  
\end{equation} 
  
(ii) ${\cal C}$ - the conventional meson wavefunction, 
$$ 
{\cal C} = \psi^{(0)}_{nlm}(\vec{r}) \; \chi_0(a_1) \chi_0(a_2) 
$$ 
  
(iii) ${\cal H}^*$ - the hybrid wavefunction (complex conjugate), 
$$ 
{\cal H}^* = \psi^{(\pm)*}_{nlm}(\vec{r})\; \beta_1 \left(a_1 \mp i 
  a_2 \right)\; \chi^*_0(a_1) \chi^*_0(a_2) 
$$ 
(where the flux tube is excited into a state with polarisation $\pm 
1$ along its axis). 
  
To be specific, we consider the transition between the unexcited tube 
and a tube that has polarisation $\pm$ along its axis (the ``body 
axis"), that axis in turn being oriented at some angles $\theta,\phi$ 
in the laboratory (the ``fixed axes"). 
   
Later we will consider both vector and axial currents in various 
multipoles. In general, to have a transition between a normal meson 
and a hybrid, one power of $\vec{a}$ will be needed in the transition 
operator. The factors multiplying $\vec{a}$ will depend on the tensor 
structure of the current (e.g. vector or axial, transverse or 
longitudinal), explicit forms being given in Appendix \ref{hybVA}. Having 
identified the presence of $\vec{a}$ we need to be able to compute its 
expectation value. This is outlined in Appendix \ref{hybtrans} for a 
``reference" operator ${\cal O}_{\mathrm{ref}} \equiv \vec{a} \cdot 
\vec{x}_i$. The $E1$ case then follows immediately.

The integral over flux-tube variables gives 
\begin{eqnarray}   
\la \chi_1, \pm| \vec{a}\cdot \vec{x}_{\pm}|\chi_0 \ra &=&  
  \pm \frac{1}{\beta_1}\left( \delta^+{\cal D}^{(1)*}_{\pm +} - \delta^-{\cal D}^{(1)*}_{\pm -} \right) \label{axplus} \\  
\la \chi_1, \pm| \vec{a}\cdot \hat{z}|\chi_0 \ra &=&  
  -\frac{1}{\sqrt{2}\beta_1}\left( \delta^+{\cal D}^{(1)*}_{0+} - \delta^-{\cal D}^{(1)*}_{0-} \right) \label{az}   
\end{eqnarray}

and so, for transition from a conventional $^1S_0$ state, for example, 
we have \be \la \mathrm{hyb}; \pm, m' | \exp 
\left[-i\vec{q}\cdot\vec{r} \frac{\mu}{m_Q} \right]\; \vec{a} \cdot 
\vec{x}_- |\mathrm{conv}; l=0 \ra = \nonumber \ee \be \int\! 
d^3\vec{r}\; \left[R_{\mathrm{hyb}}(r)\sqrt{\frac{3}{4\pi}} {\cal 
    D}^{(1)*}_{m',\pm} \right]^* e^{-i\vec{q}\cdot\vec{r} 
  \frac{\mu}{m_Q}} \left(-\frac{1}{\beta_1}\right)\left( \delta^+{\cal 
    D}^{(1)*}_{-+} - \delta^-{\cal D}^{(1)*}_{--}\right) \left[ 
  R_{\mathrm{conv}}(r) \frac{1}{\sqrt{4\pi}}\right]. 
\label{1s0me}  
\ee First expand the exponential in terms of partial wave angular 
states, contract together the three ${\cal D}$ functions and integrate 
$\int d\Omega$, which gives for the matrix element (see Appendix \ref{hybtrans}) 
   
 \[   
 -\frac{1}{\beta_1}\frac{1}{\sqrt 3} \delta_{m',-1}\left[\left(_f{\la 
       j_0 \ra}_i - \frac{1}{2} {_f \la j_2 \ra_i}\right) 
   \left(\delta^+ - \delta^-\right) - i\frac{3}{2}{_f \la j_1 \ra_i} 
   \left(\delta^+ + \delta^-\right) \right] 
 \]   
 where we now only need to calculate the radial expectation values of 
 the spherical Bessel functions. 
   
    
 In the previous equations the factors $\delta^\pm$ refer to the flux 
 tube polarisation transverse to the body vector $\vec{r}$, while the 
 $\delta_{m',\pm 1}$ refers to the meson's total angular momentum 
 projection in the fixed axes $(\hat{x},\hat{y},\hat{z})$ The parity 
 eigenstates in the flux tube are given in ref.\cite{IP}. They are 
 linear superpositions of states where the flux-tube has polarisation 
 $\pm 1$. Following that reference we denote the number of positive or 
 negative helicity phonon modes by $\{n_+,n_-\}$, which for our 
 present purposes will be $\{1,0\}$ or $\{0,1\}$. Parity eigenstates 
 are then the linear superpositions 
\[   
|{\cal P}= \pm \rangle \equiv \frac{1}{\sqrt 2} \left( |\{1,0\}\rangle 
  \mp |\{0,1\} \rangle \right) 
\]   
The effect is that when we take expectation values for parity 
eigenstates, the terms proportional to $(\delta^+ \pm \delta^-)$ will 
be destroyed for the ``wrong" parity, and amplified by $\sqrt{2}$ for 
the ``correct" parity. 
    
This is the source of the extra overall factor of $\sqrt{2}$ in the 
following expressions for transitions to specific parity eigenstates. 
With this preamble we now proceed to complete the expression for 
parity eigenstates. 
   
In the above the argument of Bessel functions is $j_n(-|\vec{q}|r 
\frac{\mu}{m_Q})$. Using $j_{0,2}(-x)\equiv j_{0,2}(x)$ and 
$j_1(-x)\equiv - j_1(x)$ we can replace the argument of the Bessel 
functions to be $j_n(|\vec{q}|r \frac{\mu}{m_Q})$ and gather together 
the general structure of the matrix elements for the various 
polarisation states 
\[   
\la \mathrm{hyb}; \pm, m' | \exp \left[-i\vec{q}\cdot\vec{r} 
  \frac{\mu}{m_Q}\right]\; \vec{a} \cdot \vec{x}_{\pm} |\mathrm{conv}; 
l=0 \ra 
 \]   
 \be = \pm \frac{1}{\beta_1}\sqrt{\frac{2}{3}} \left\{ \delta({\cal 
     P}=+)\left({_f\la j_0\ra_i} - \frac{1}{2}{_f\la j_2\ra_i}\right) 
   \mp \frac{3i}{2} \delta ({\cal P}=-) {_f\la j_1 \ra_i} \right\} 
 \delta_{m',\pm 1} 
\label{xplus}   
\ee  
and 
 \[   
 \la \mathrm{hyb}; \pm, m' | \exp \left[-i\vec{q}\cdot\vec{r} 
   \frac{\mu}{m_Q}\right]\; \vec{a} \cdot \hat{z} |\mathrm{conv}; l=0 
 \ra 
\]  
\be  
= - \frac{1}{\sqrt{2}\beta_1}\sqrt{\frac{2}{3}} \left\{ 
  \delta({\cal P}=+)\left({_f\la j_0\ra_i} + {_f\la j_2\ra_i}\right) 
\right \} \delta_{m',0} 
\label{xz}   
\ee  
These are the main equations that set the normalisation for hybrid 
excitation involving any operator.  To illustrate how they are applied 
we return to the specific example of $E1$ electromagnetic transitions.

The $E1$ operator is eqn.(\ref{opE1}), thus we simply rescale the 
above expressions accordingly. Noting that 
$|\vec{\epsilon}_\pm|/|\vec{x}_\pm| = 1/\sqrt 2$ we have effectively 
\[   
{\cal O}_{E1}/{\cal O}_{\mathrm{ref}} = -|\vec{q}| \left( 
  \frac{e_Q}{m_Q} + \frac{e_d}{m_d} \right) \sqrt{\frac{b}{\pi^3}} 
\beta_1 r 
\]   
The matrix element for the transition from $^1S_0$ to spin -singlet 
$\mathrm{hybrid}(1^{++})$ then follows upon multiplying this ratio by 
the reference form (\ref{xplus}) (for ${\cal P}=+$), in the $|\vec{q}| 
r \ll 1$ limit 
\[   
\la \mathrm{hyb}; \pm, m' | \vec{a} \cdot \vec{x}_+ |\mathrm{conv}; 
l=0 \ra = + \frac{1}{\beta_1}\sqrt{\frac{2}{3}} \delta({\cal P}=+) 
\delta_{m',+1} 
\]   
giving finally 
 \[   
 {\cal M}( \mathrm{conv}(0^{-+}) \; \gamma \rightleftharpoons 
 \mathrm{hyb}(1^{++})) = \left(\frac{e_1}{m_1} + 
   \frac{e_2}{m_2}\right) |\vec{q}| \sqrt{\frac{2b}{3\pi^3}} 
 \delta_{m',+1} \int \!r^2 dr\; R^*_{\mathrm{hyb}}(r)\; r \; 
 R_{\mathrm{conv}}(r) 
  \]   
  which is the form used in ref.\cite{cd03}. 
   
  This applies immediately to the excitation of the hybrid 
  $a_{1H}^{\pm}$ in $\gamma \pi^{\pm} \to a_{1H}^{\pm}$ where there is 
  no spin flip between the spin singlets $\pi$ and $a_{1H}$. Note that 
  this transition requires charged states, the neutral modes vanishing 
  in accord with charge conjugation.   
\be  
{\cal M}(\gamma \pi 
  \rightleftharpoons a_{1H}) \approx \left(\frac{e_1 + 
      e_2}{m_n}\right) {_H\la r \ra_\pi} |\vec{q}_H| \sqrt{\frac{2 
      b}{3\pi^3}} 
  \label{a1hmatrix}   
  \ee  
\be  
{\cal M} (\gamma \pi \rightleftharpoons b_1) \approx 
  \left(\frac{e_1 - e_2}{m_n}\right) {_b \la r \ra_\pi} |\vec{q}_b| 
  \frac{m_n}{2 \sqrt{3}} 
  \label{b1matrix}  
  \ee  
So the ratio of widths becomes  
\be \frac{\Gamma_{E1}(a_{1 H}^+ \to \pi^+ 
  \gamma)} {\Gamma_{E1}(b_{1 Q}^+ \to \pi^+ \gamma)} = 
\frac{72}{\pi^3} \frac{b}{m_n^2} \left | \frac{_{a_H}\la r 
    \ra_\pi}{_b\la r \ra_\pi} \right |^2 \left [ \frac{|\vec{q}_H|^3 
    \exp \left( - \frac{|\vec{q}_H|^2}{8 \bar \beta_H^2} \right)} 
  {|\vec{q}_b|^3 \exp \left(- \frac{|\vec{q}_b|^2}{8 \bar \beta_b^2} 
    \right) } \right ] 
\label{ratecompare}  
\ee  
where the factor in square brackets includes the $q^3$ phase-space 
and a ``typical'' form-factor taken from the case of 
harmonic-oscillator binding \cite{CDK02}. These factors model the 
non-trivial hybrid meson mass dependence of the width\footnote{Keeping 
  the full spherical Bessel functions in the radial overlap gives a 
  form-factor very similar to this. Even calculated true to the model, 
  the form-factors should be considered to be at best rough guides to 
  the mass dependence especially in regions where the non-rel approximation is failing}.

Compare the form of the ratio of $E1$ widths, (after removing a factor 
of 9 due to the different charge factors), with the transverse 
contribution to the elastic charge radius, eqn.(\ref{r2}).  In the 
approximations that we have used here, the E1 transitions to the 
leading states saturate the dipole sum rule. 
   
Our calculation of the relative strengths of the matrix elements for 
hybrid and conventional $E1$ transitions, eqns.(\ref{a1hmatrix} ,\ref{b1matrix}), in the flux-tube model, suggests a way of calculating 
this more directly in Lattice QCD. The essential features of the 
electromagnetic matrix elements are 
   
(i) initial charged $\pi^{\pm}$; 
   
(ii) $E1$ transition operator with no spin-flip; 
   
(iii) compute matrix element to $J^P = 1^+$ final states with G-parity 
= $\pm 1$. 
   
The transition to $G$-parity +1 is to the conventional axial meson, 
while that to $G=-1$ is only accessible for the hybrid configuration. If 
further one calculated $\la r^2 \ra$ for the $\pi^{\pm}$ one could 
assess how well the sum rule is saturated by these states, quantify 
the ``penalty" for exciting the gluonic modes or hybrids in general 
and potentially assess the role of such configurations in the $\pi$ 
wavefunction.

\subsection{E1 Rates}   
   
In the Isgur-Paton adiabatic model\cite{IP} with a variational  
harmonic-oscillator solution\footnote{see Appendix \ref{hybham}} we obtain $ |  
_H\la r \ra_\pi / _b\la r \ra_\pi |^2 \approx 1.0$, so the radial  
moments do not suppress hybrid production.  We follow ref.\cite{IP}  
and use the standard parameters $b=0.18 \mathrm{GeV}^2, m_n=0.33  
\mathrm{GeV}$ so that the prefactor $\frac{72}{\pi^3} \frac{b}{m_n^2}  
\approx 3.8$ and hence there is no hybrid suppression from the  
flux-tube dynamics.  
   
\begin{figure} 
\begin{center} 
  \psfragscanon 
  \psfrag{XX}{$m(1^{++}_H) / \mathrm{MeV}$} 
  \psfrag{Y}{$\Gamma/\mathrm{keV}$} 
 \includegraphics[width=3in]{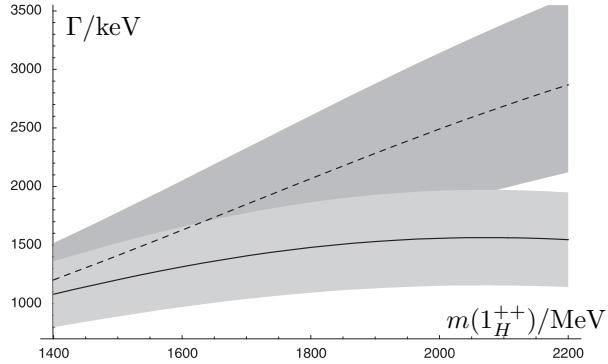}
\caption{E1 width as a function of $1^{++}$ hybrid mass.     
                The solid line is for $\beta_\pi = 335 \mathrm{MeV}$. The     
                dashed line is for  $\beta_\pi = 540 \mathrm{MeV}$ \cite{SWANSON}.     
                The shaded grey areas are the uncertainties due to the error on the     
                experimental rate used as normalisation. \label{figE1}} 
\end{center} 
\end{figure}

Within our variational solution $\beta_H = 255 \mathrm{MeV}, \beta_b =  
281 \mathrm{MeV}, \beta_\pi = 335 \mathrm{MeV}$, so we see the $p=1$  
hybrid state being roughly the same size as the $L=1$ conventional  
state. The main uncertainty is the computed size of the  
$\pi$\cite{CDK02}.  Assuming that this hybrid has mass $\sim 1.9  
\mathrm{GeV}$\cite{hybrids,IP,bcs}, and using the measured width  
$\Gamma(b_{1}^+ \to \pi^+ \gamma) = 230 \pm 60 \mathrm{keV}$\cite{PDG}  
we predict that  
\be   
\Gamma(a_{1 H}^+ \to \pi^+ \gamma) = 2.1 \pm 0.9 \mathrm{MeV}.  
\label{a1Hrate} 
\ee  
where the error allows for the uncertainty in  
$\beta_{\pi}$\cite{CDK02, SWANSON}.  
We present in Figure.(\ref{figE1}) the dependence of the radiative width on the mass of the   
hybrid.   
   
  
The equivalent $E1$ process for $S=1$ is $b_{JH} \to \rho  
\gamma$, where the only difference from the $S=0$ case is the addition  
of $L,S$ Clebsch-Gordan factors coupling the $Q\bar{Q}$ spin and flux-tube angular momentum to the total $J$ of the hybrid meson in  
question. As above, the charge conjugation of the initial and final states is the same, thus  
 $\Delta C =0$, and  
the amplitude is proportional to $e_1 + e_2$. Consider absorption of a  
positive helicity photon. The hybrid state is constructed as follows   
\[   
| J^{+-}, m_J \rangle = \sum_{m_S,m'} \langle 1m';1m_S | J m_J \rangle  
\frac{1}{\sqrt{2}}\left \{ |\mathrm{hyb}; +, m' \ra - |\mathrm{hyb};  
  -, m'\ra\right \} |S=1, m_S \ra  
\]   
and the matrix element becomes  
  \[   
  {\cal M}(\gamma \rho \rightleftharpoons b_{J H})= \langle 1 +\!\!1 ;  
  1 m_{\rho} | J m_J \rangle \left(\frac{e_1 +  
      e_2}{m_n}\right) {_H\langle r \rangle_\rho} |\vec{q}_H|  
  \sqrt{\frac{2 b}{3\pi^3}}  
  \]    
  We find for $J=0,1,2$ in this $E1$ limit, normalising against measured $\Gamma(f_1 \to \rho \gamma)$,  
\be   
\Gamma(b_{J H}^+ \to \rho^+ \gamma) = 2.3 \pm 0.8 \mathrm{MeV}.  
\label{bJHrate} 
\ee  
where the error reflects the uncertainties in the conventional $E1$  
strength and $\beta_{f_1}$ and where we have taken $m_H =  
1.9\mathrm{GeV}$.  
  
We present in Table \ref{supertable}, formulae for $E1$ radiative matrix elements between conventional and hybrid states. Of particular interest is the rate of production of the isovector $1^{-+}$($\pi_{1H}$) at 1.6 GeV. We use the $E1$ decay of this state to $a_2$ as an explicit example of the use of Table. \ref{supertable}. With a positive helicity photon there are three helicity amplitudes, corresponding to $m_J=-2,-1,0$,
\begin{eqnarray}
  {\cal M}_{-2} &=& \sqrt{\frac{3}{4}}   \la 1-1;1-1|2-2\ra \la10;1-1|1-1\ra {\cal M} = \frac{1}{\sqrt{2}}  \sqrt{\frac{3}{4}} {\cal M} \nonumber \\
  {\cal M}_{-1} &=& \sqrt{\frac{3}{4}} \left(   \la 10;1-1|2-1\ra \la1+1;1-1|10\ra + \la 1-1;10|2-1\ra \la10;10|10\ra \right){\cal M} = \frac{1}{2}  \sqrt{\frac{3}{4}} {\cal M} \nonumber \\
  {\cal M}_{0} &=& \sqrt{\frac{3}{4}} \left(   \la 1-1;1+1|20\ra \la10;1+1|1+1\ra + \la 10;10|20\ra \la1+1;10|1+1\ra \right){\cal M} = \frac{1}{2\sqrt{3}}  \sqrt{\frac{3}{4}} {\cal M},\nonumber
\label{helamp}
\end{eqnarray}
where ${\cal M} = |\vec{q}| \; 87 \times 10^{-3} \mathrm{GeV}^{-1}$. Using eqn. \ref{width} we obtain
\begin{eqnarray}
\Gamma_{E1}(\pi_{1H} \to a_2 \gamma) &=& 4 \frac{E_{a_2}}{m_{\pi_{1H}}} |\vec{q}| \frac{1}{3} \left( |{\cal M}_{-2}|^2 +|{\cal M}_{-1}|^2 +|{\cal M}_{0}|^2 \right) \nonumber \\
&=& \frac{E_{a_2}}{m_{\pi_{1H}}} |\vec{q}|^3 \left( \frac{1}{2} + \frac{1}{4} + \frac{1}{12} \right) (87 \times  10^{-3} \mathrm{GeV}^{-1})^2, \nonumber
\end{eqnarray}
so that for a $\pi_{1H}$ at 1.6 GeV the width is $\sim 90 \mathrm{keV}$. Given that $a_2$ exchange is suppressed relative to $f_2$ in photoproduction, this is unlikely to be a major production route in $\gamma p \to \pi_{1H} n$. Pion exchange provides an opportunity via the $M1$ multipole but this goes beyond our present discussion. If calculated via VDM from the $\pi_{1H} \to \pi \rho$ rate, where the photon-rho flux-tube is excited by a ``pion current'' we find \cite{cdpirho} $\Gamma(\pi_{1H} \to \gamma \pi) \sim 50-200 \mathrm{keV}$ which is similar to that found in \cite{pageanas}.

A state potentially interesting in heavy-flavour decay (see next section) is the axial hybrid Kaon, we find that this state has an $E1$ width to $K \gamma$ of $300 - 1000 \mathrm{keV}$ (assuming mass $\sim 2 \mathrm{GeV}$). This state could be seen in photoproduction by looking in the $K \pi \pi \Lambda$ end-state.

\begin{table}   
\begin{center}
\begin{tabular}{c| c| c| c }   
state & & $u\bar{d}$ & $u\bar{s}$  \\   
\hline   
$^1 S_0$ & $\times 1$ & $\gamma \pi^+ \to a_{1 H}^+$ & $\gamma K^+ \to  K_{1A H}^+$  \\   
& & 56 (23) & 43 (23) \\    
 $^3 S_1$ & $\times \la 11;1m_i | J_H m_H \ra$ & $\gamma \rho^+ \to b_{J H}^+$ & $\gamma K^{*+} \to  K_{J B H}^+$  \\   
& & 56 & 43  \\   
$^1 P_1$ & $\times \sqrt{\frac{3}{2}} \la 11;1m_i | 1 m_H \ra $ &  $\gamma b_1^+ \to \rho_H^+$ & $\gamma K_{1B}^+ \to  K_H^{*+}$ \\   
& & 87 & 68 \\   
$^3 P_J$ & $\times \displaystyle{\sum_{m_L,m_S} \la 1 m_L;1 m_S | J m_J \ra} $ & $\gamma a_{J}^+ \to \pi_{J_H H}^+$ & $\gamma K_{JA}^+ \to  K_{J_H H}^+$ \\   
& $\cdot \sqrt{\frac{3}{4}} \la 1 m_L+1;1 m_S | J_H m_H \ra^* $ & 87 & 68 \\   
\hline   
\end{tabular}   
\caption{Photon-Meson-Hybrid $E1$ matrix elements: ${\cal M} = \left( \frac{e_1}{m_1} + \frac{e_2}{m_2} \right) | \vec{q} | \sqrt{\frac{2 b}{3 \pi^3}} {_H \la r \ra_i}$ should be multiplied by the Clebsch-Gordan factor in the second column to give the overall matrix element for a positive helicity photon. The numbers quoted in columns three and four are ${\cal{M}} /|\vec{q}| \; (10^{-3} \mathrm{GeV}^{-1}) $, evaluated using the results of Appendix \ref{hybham}, except those   
in brackets which use the $\beta$-values of \cite{SWANSON}.    
\label{supertable}    
}
\end{center}   
\end{table}   
   
Note that these $E1$ transitions are only possible with charge  
exchange and so cannot occur between flavourless states. In particular  
they are absent for $c\bar{c}$ and $b\bar{b}$. Thus, for example, the  
transitions $\psi(3685) \to \gamma \chi_J$ can receive no contribution from any  
hybrid component of the $\psi(3685)$ wavefunction (assuming here that the $\chi_{JH}$ 
states are $\gg 4$ GeV in mass and so do not mix measurably into the \cc ~ states.)  
   
This illustrates the principal for calculating matrix elements for  
hybrid excitation by currents.  We can extend this now to vector and  
axial currents of arbitrary polarisation and apply to the production of hybrid mesons  
in the weak decays of heavy flavours.  This requires a  
non-relativistic decomposition of the currents (Appendix \ref{hybVA}), then we identify the  
terms linear in $\vec{a}$ which can cause transitions between  
conventional and first excited hybrids.  Finally all one then needs to  
do is to read off the relevant operator, scale by the reference  
operator as above, and determine the relevant matrix elements.  
    
\section{Heavy Flavour Decays to Hybrid Mesons} 
 
In the first half of this paper we have described the method of 
calculation used to compute amplitudes for processes involving a hybrid meson, a 
conventional meson and a current by considering the specific example 
of hybrid meson radiative decay. The results obtained are applicable 
to the forthcoming hybrid photoproduction experiments at 
Jefferson Lab \cite{jlab}. 
 
We will now use the same techniques to calculate the rate of 
production of hybrids in heavy flavour decays. Such a calculation is 
timely in view of the orders of magnitude increase in statistics 
 on a wide range of exclusive decay 
channels anticipated 
at present and upgraded B and charm factories. 
We will consider in particular the supposed excess of events 
in the inclusive decay $B \to J/\psi X$ at low $J/\psi$-momentum\cite{psidata} as 
this may have an explanation in terms of hybrid Kaon production. 
 
In Tables (\ref{Dpitab}) and (\ref{BpsiK}) we identify the exclusive channels in which hybrids may 
be clearly observed. We begin by outlining the model used to describe 
the non-leptonic weak decay process. 
 
\subsection{Naive Factorisation Model} 
 
The matrix element for decays $B \to M_1M_2$ will be written in the 
generic form (see e.g. Ref.(\cite{NeuSte})), 
\begin{equation}  
\la M_1M_2 | {\cal H}_{\mathrm{eff}} | B \ra = 
\frac{G_F}{\sqrt{2}} V_{q'_1b}V_{q'_2q_2} {\cal O} 
 \label{weakme}  
\end{equation} 
with 
\begin{equation}  
{\cal O} = a_1(\mu) \la M_1|\bar{q'}_1(V-A)b| B \ra \la 
 M_2 | \bar{q}_2(V-A)q'_2 | 0 \ra + a_2(\mu) \la 
 M_2|\bar{q}_2(V-A)b|B\ra \la M_1 | \bar{q'}_1(V-A)q'_2 | 0 \ra. 
  \label{factorop}  
\end{equation} 
The model arises from performing the QCD renormalisation of the weak 
interaction, supposing that one of the mesons is created from the 
vacuum by a current with the same quantum numbers.  
Final state interactions are ignored. $a_1$, $a_2$ are considered 
as 
phenomenological parameters determined by fitting to known 
conventional decay rates and not as the 
Wilson coefficients they would be in the strict theory. 
 
Decays proceeding by the first term only are labeled ``Class I'', an 
example being $\bar{B}^0 \to \pi^- D^{(*)+} $ which in this model has 
amplitude, 
\begin{equation} 
\frac{G_F}{\sqrt{2}} V_{bc}V_{ud} a_1 \la \pi^- | \bar{d} A_\mu u | 0 \ra  
     \la D^{(*)+} | \bar{c} (V^\mu - A^\mu) b | \bar{B}^0 \ra. 
  \label{class1} 
\end{equation}  
The pion creation current is conventionally parameterised by $ \la 
\pi^-(q) | \bar{d} A_\mu u | 0 \ra \equiv i f_\pi q_\mu$ and we 
compute the hadronic matrix element using our non-relativistic 
quark-flux-tube model. 
 
Decays proceeding only by the second term in (\ref{factorop}) are 
labeled ``Class II'' or ``color-suppressed'' and include the well 
tested channel $B^+ \to K^{(*)+} J/\psi$ with amplitude 
\begin{equation} 
\frac{G_F}{\sqrt{2}} V_{bc}V_{cs} a_2 \la J/\psi | \bar{c} V_\mu c | 0 \ra  
     \la K^{(*)+} | \bar{s} (V^\mu - A^\mu) b | B^+ \ra, 
  \label{class1} 
\end{equation}  
where the $J/\psi$ current has Lorentz decomposition $\la J/\psi(q, 
\epsilon) | \bar{c} V_\mu c | 0 \ra \equiv \epsilon_\mu^* f_\psi 
m_\psi$. 
  
Models of this type are more thoroughly discussed in \cite{NeuSte} 
where they are demonstrated to successfully 
predict weak decay rates of the $B$-meson  to a wide range of 
conventional exclusive channels. 
 
We use meson state normalisations as in \cite{isgw} such that a factor  
$\sqrt{4m_B m_M}$ appears in all hadronic matrix elements, 
and where $M$ is the meson not created from the vacuum.  
With this normalisation the decay widths are written in the following manner, 
\begin{equation}  
\Gamma(\bar{B}^0 \to \rho^- D^{+(*)}) = \frac{G_F^2}{16 \pi} \frac{|\vec{q}|}{m_B^2} |V_{bc} V_{ud}|^2 |a_1|^2 \sum_{\mathrm{helicities}} |{\cal M}|^2 
\label{pwidth}  
\end{equation} 
where ${\cal M} = \la \rho^- | \bar{d} V_\mu u | 0 \ra \la D^{(*)+}  
| \bar{c} (V^\mu - A^\mu) b | \bar{B}^0 \ra$ and where  
to be specific we have shown the case of the Class I decay to a conventional vector.  
We will use $a_1=1.05$, $a_2=0.25$ for all $B$-decays, these numbers are compatible 
 with those in \cite{NeuSte} and give a good fit to the conventional decays. 
 
\subsection{Application to $B$ decays producing Hybrid mesons} 
 
In this section we extend the established model described above to 
include hybrids in the end state, in much the same was as we extended 
the conventional radiative decay formalism previously. We simply 
include the additional flux-tube transverse degree of freedom, 
$\vec{a}$, in the currents and wavefunctions. As in the radiative 
case this both modifies conventional decay rates (by changing the 
charge radius or Isgur-Wise function as detailed in Section \ref{model}) and allows for hybrid excitation. 
 
Previous models of hybrid production in this type of process 
\cite{slowJ}, \cite{petrov} have assumed that hybrids are created by 
a color-octet current with some undetermined strength; we differ 
somewhat in that the flux-tube model has had color degrees of freedom 
removed from the beginning. We have the hybrid produced by the same 
current that produces the conventional states. Rather than  
color-octet and color-singlet factors controlling the production of hybrids  
and conventional states, we have 
the hybrid production rate determined by excited flux-tube mode in contrast 
to ground state 
flux-tube mode for conventional states. 
The relative rate for hybrid production generically is then 
a factor ${\cal O}(b/m^2)$. 

When considering weak decays of heavy-light systems we use the following form of the quark position vector,
\[   
\vec{r}_{Q(d)} = \vec{R} \pm \vec{r}\frac{\mu}{m_{Q(d)}} + \vec{a} 
\frac{2 \beta_1 r}{m_Q+m_d} \sqrt{\frac{2b}{\pi^3}}. 
\] 
This differs slightly from the form used previously, eqn.(\ref{isgr}). It has the advantage that under a finite change in heavy quark mass (i.e. a flavour changing transition) $\vec{r}$, which here is the interquark separation vector, and $\vec{a}_p$ are unchanged. For details see \cite{jd03b}. 
 
\subsection{$B$ decays to Hybrid and conventional pseudoscalar} 
 
The matrix element for such a decay is 
\[ 
{\cal M}(B \to P H) = i f_P q \cdot \la H |(V-A)| B \ra. 
\] 
Examples include $\bar{B}^0 \to \pi^- D_H^+$ (Class I) and $\bar{B}^0 
\to D^0 n\bar{n}_H$ (Class II)\footnote{ recent theoretical work 
  \cite{Stewart} suggests that naive factorisation can be a poor 
  approximation for processes $\bar{B}^0 \to D^0 +$ light hadrons}. 
Thus to compute the branching ratio for $\bar{B}^0 \to \pi^- D_H^+$ we 
need the values of the operators $q\cdot A$ and $q \cdot V$ between 
$B$ and $D_H$ wavefunctions. In Appendix \ref{hybVA} we compute the terms in 
the non-relativistic reduction of $V_\mu$ and $A_\mu$ that are linear 
in $\vec{a}$, and which therefore can hence induce transitions between conventional and 
hybrid mesons. 
 
We find, for the maximally parity violating current, 
\[ 
q \cdot{V}_H = e^{-i\vec{q}\cdot 
  \vec{r}\frac{m_d}{m_D}}\sqrt{\frac{2b}{\pi^3}}\beta_1|\vec{q}| 
\times 
\] 
\begin{equation}   
\left\{i \vec{a}\cdot \hat{z} \left[  
\frac{\pi}{2 m_b} \left(1 + \frac{m_b}{m_c} + \frac{|\vec{q}|}{2m_c} \right) + |\vec{q}| \frac{2r}{m_D}  
\left(1+ \frac{|\vec{q}|}{2m_c}  \right) \right] +\vec{\sigma}\cdot \hat{z} \wedge \vec{a}\frac{\pi}{2 m_b}  
\left( -1 + \frac{m_b}{m_c} - \frac{|\vec{q}|}{2m_c}\right) \right\}  
\label{qvh}  
\end{equation}   
and for the parity conserving current, 
\begin{equation}  
q \cdot{A}_H = e^{-i\vec{q}\cdot \vec{r}\frac{m_d}{m_D}}\sqrt{\frac{2b}{\pi^3}}\beta_1|\vec{q}| \left\{-i \vec{\sigma}\cdot \vec{a}  
\frac{\pi}{2 m_b} \left(1 + \frac{m_b}{m_c} + \frac{|\vec{q}|}{2m_c} \right)  
- i |\vec{q}| \sigma_z \vec{a}\cdot \hat{z} \frac{2r}{m_D}\left(1 + \frac{|\vec{q}|}{2m_c}  \right) \right\}  
\label{qdota}  
\end{equation}  
having approximated $q^2=m_\pi^2 \approx 0$, as appropriate for 
$B$ decays. 
 
\subsubsection*{Spin-singlet Hybrids, $D_H (1^{\pm(\pm)})$} 
 
Since $\la \vec{\sigma} \ra = 0$ for transition to a hybrid 
 where the \qq~ spins are coupled to $S=0$, the parity conserving current eqn.(\ref{qdota}) 
 plays no role here. Non-vanishing contributions come entirely from 
  the first term of the parity violating current, eqn.(\ref{qvh}). We thus have a matrix element, 
\[ 
{\cal M}(\bar{B}^0 \to \pi^- D_H^+(S=0)) = i f_\pi \times 
\] 
\[ 
 \la D_H; S=0 | e^{-i\vec{q}\cdot \vec{r}\frac{m_d}{m_D}}\sqrt{\frac{2b}{\pi^3}}\beta_1|\vec{q}| \vec{a}\cdot \hat{z} \left[  
\frac{\pi}{2m_b} \left(1 + \frac{m_b}{m_c} + \frac{|\vec{q}|}{2m_c} \right) + |\vec{q}| \frac{2r}{m_D}  
\left(1+ \frac{|\vec{q}|}{2m_c}  \right) \right] | B; \substack{S=0 \\L=0} \ra. 
\] 
The integrals over $\vec{a}$ and angles $\theta, \phi$ are performed in  
Appendix \ref{hybtrans}, and we take the result eqn.(\ref{xz}), 
 \[  
\la \mathrm{hyb}; {\cal P}=\pm, m' | \exp \left[-i\vec{q}\cdot\vec{r} 
   \frac{\mu}{m_Q}\right]\; \vec{a} \cdot \hat{z} 
 |\mathrm{conv}; l=0 \ra  
\] 
\[  
= - \frac{1}{\sqrt{2}\beta_1}\sqrt{\frac{2}{3}} 
   \delta({\cal P}=+) \left( \la j_0\ra + \la j_2\ra \right) 
 \delta_{m',0}. 
\] 
where in the current example the argument of the spherical Bessel functions  
is $|\vec{q}| r \frac{m_d}{m_d+m_c}$. Thus we have the matrix elements, 
\begin{eqnarray} 
{\cal M}(\bar{B}^0 \to \pi^- D_H^+(1^{-(-)})) &=& 0,\\ 
{\cal M}(\bar{B}^0 \to \pi^- D_H^+(1^{+(+)})) &=& - i f_{\pi} \sqrt{4m_Bm_D} 
\sqrt{\frac{2b}{3\pi^3}}|\vec{q}| \times \nonumber \\ 
&&\hspace{-1 cm} \left \{ \frac{\pi}{2 m_b} \left(1 + \frac{m_b}{m_c} + \frac{|\vec{q}|}{2m_c} \right) (\la j_0\ra + \la j_2\ra) + \frac{6}{m_d}\left(1+ \frac{|\vec{q}|}{2m_c} \right) \la j_1\ra \right \}.  
\end{eqnarray}  
That the $1^{-(-)}$ amplitude is zero was to  
be expected as one cannot maximally violate parity and 
 conserve angular momentum in a process $0^- \to 1^- 0^-$  
 in any partial wave, whereas the $1^{+(+)}$ amplitude is  
 non-zero as $0^- \to 1^+ 0^-$ in a $P$-wave respects the symmetries. 
 
\subsubsection*{Spin-triplet Hybrids, $D_H (J^{\pm(\mp)})$} 
 
To excite spin-triplet hybrids we require operators linear in 
$\vec{\sigma}$ and $\vec{a}$, which feature in both $q \cdot A$ and $q 
\cdot V$, eqns.(\ref{qvh},\ref{qdota}). 
 
For simplicity in presentation we shall define $\rho_{\pm} \equiv \frac{\pi}{2m_b} \left(1 \pm \frac{m_b}{m_c} + \frac{|\vec{q}|}{2m_c} \right)$. We can decompose $\vec{\sigma} \cdot \hat z \wedge \vec{a} = 
 \frac{1}{2i}(\sigma_+ \vec{a} \cdot \vec{x}_- -\sigma_- \vec{a} \cdot 
 \vec{x}_+ )$ and $\vec{\sigma} \cdot \vec{a} = \frac{1}{2}(\sigma_+ 
 \vec{a} \cdot \vec{x}_- +\sigma_- \vec{a} \cdot \vec{x}_+ ) + 
 \sigma_z \vec{a}\cdot \hat z$, where $\sigma_\pm, \sigma_z$ are 
 normalised such that $\la S=1, m_S|\substack{\sigma_\pm 
   \\\sigma_z}|S=0\ra = \left[\substack{\mp \sqrt{2} \delta(m_S,\pm 1)\\ 
   \delta(m_S,0)}\right]$. 
  
 The $S=1$ of the quarks and $L=1$ of the flux tube combine together 
 to give the $J$ of the hybrid meson, with the appropriate 
 Clebsch-Gordan coefficient $\la 1m_L;1m_S | Jm_J \ra$. Performing the integrations over $\vec{a}$ and angles $\theta, \phi$ as in Appendix \ref{hybtrans} gives for the parity violating current, 
 \begin{eqnarray} 
 {\cal M}_{q \cdot V} = 
 - f_\pi \sqrt{4m_Bm_D} \sqrt{\frac{2b}{3 \pi^3}} |\vec{q}| \rho_-   
  \sum_{m_L,m_S}\la 1m_L;1m_S | Jm_J \ra \times \nonumber \\ 
    \left \{\begin{array}{l} 
    \delta({\cal P}=+)\left(\la j_0\ra - \frac{1}{2}\la j_2\ra \right)  
 \left(\delta_{m_L,-1}\delta_{m_S,+1}  - \delta_{m_L,+1}\delta_{m_S,-1}\right)  
       \\  
 + \delta ({\cal P}=-) \frac{3i}{2} \la j_1 \ra   
 \left(\delta_{m_L,-1}\delta_{m_S,+1}  
   + \delta_{m_L,+1}\delta_{m_S,-1}\right)  
 \end{array} \right \}. 
 \label{mva2}  
\end{eqnarray} 
The pattern of amplitudes for various $J^{PC}$ follows from the combination of Clebsch-Gordan coefficients, 
 \[  
 \delta({\cal P}=+)(\la j_0\ra - \frac{1}{2}\la j_2\ra)  
 \left[\la 1-1;1+1|J0 \ra - \la 1+1;1-1| J0 \ra   \right]  
 \]  
 \[  
+ \delta ({\cal P}=-)\frac{3i}{2}  \la j_1 \ra   
 \left[\la 1-1;1+1|J0 \ra + \la 1+1;1-1| J0 \ra   \right]  
 \]  
 
 Thus we find, for the maximally parity violating current, 
 \[  
 {\cal M}_{q \cdot V}\left(\begin{array}{c} 
     0^{+(-)}\\ 
     1^{-(+)}\\ 
     2^{+(-)} 
                   \end{array} 
                 \right) = 0 
 \]  
 \begin{equation}  
 {\cal M}_{q \cdot V}\left(\begin{array}{c} 
                     0^{-(+)}\\ 
                     1^{+(-)}\\ 
                     2^{-(+)} 
                   \end{array} 
               \right)  
= - \sqrt{4 m_Bm_D} \sqrt{\frac{2b}{3\pi^3}} f_\pi |\vec{q}| \rho_-  
 \left\{ \begin{array}{c} 
         \frac{i}{\sqrt{3}}\la j_1 \ra \\ 
         -\sqrt{2}(\la j_0 \ra - \frac{1}{2} \la j_2 \ra) \\ 
         \frac{i}{\sqrt{6}} \la j_1 \ra 
         \end{array} 
 \right\}. 
 \end{equation}  
An exactly analogous process for the parity conserving current gives, 
 \[  
 {\cal M}_{q \cdot A}\left(\begin{array}{c} 
     0^{-(+)}\\ 
     1^{+(-)}\\ 
     2^{-(+)} 
                   \end{array} 
                 \right) = 0 
 \]  
 \begin{equation}  
 {\cal M}_{q \cdot A}\left(\begin{array}{c} 
                     0^{+(-)}\\ 
                     1^{-(+)}\\ 
                     2^{+(-)} 
                   \end{array} 
               \right) =\sqrt{4m_Bm_D} \sqrt{\frac{2 b}{3\pi^3}}f_\pi |\vec{q}|  
\left( \frac{\rho_+}{\sqrt{2}} \left\{ \begin{array}{c} 
                        \sqrt{6}\la j_0 \ra \\ 
                        - 3i\la j_1  \ra \\ 
                        -\sqrt{3} \la j_2 \ra 
                        \end{array} 
                        \right\} 
+ \frac{2\sqrt{3}}{m_d} \left(1+\frac{|\vec{q}|}{2m_c}\right) \la j_1 \ra 
             \left\{ \begin{array}{c} 
                     1 \\ 
                     0 \\ 
                     - \sqrt{2} 
                     \end{array} 
             \right\} 
\right).         
\end{equation}  
As in the spin-singlet case we can understand the zero values as 
originating in the need for angular momentum conservation while 
applying the appropriate parity selection rule. 
 
\subsubsection*{Numerical Estimates} 
The model parameters and variational wavefunctions used to evaluate the radial matrix elements $\la D_H | j_L | B\ra$ are described in Appendix \ref{hybham}. We obtain the following results for the branching fractions of the $\bar{B}^0$ meson to the exclusive channel $D_H^+(J^{P(C)}) \pi^-$, presented in Table.(\ref{Dpitab}). 
\begin{table}[h] 
  \begin{center} 
  \begin{tabular}{c|ccc|c} 
    $m_{D_H}$ (GeV) & 2.7 & 3.0 & 3.3 \\ 
    $\substack{\la j_0 \ra \\ \la j_1 \ra \\\la j_2 \ra}$ & $ \substack{0.62 \\ 0.24 \\ 0.06}$ & $ \substack{0.65 \\ 0.22 \\ 0.05}$  & $ \substack{0.67 \\ 0.21 \\ 0.04}$\\ 
\hline 
    $1^{+(+)}$ &$6.1$& $4.5$&$3.0$ & $\times 10^{-4}$\\ 
    $1^{-(-)}$ &$0$ &$0$ &$0$\\ 
\hline     
    $0^{-(+)}$ & $3.1$ & $2.5$ & $1.9$ &$ \times 10^{-7}$\\ 
    $1^{+(-)}$ & $1.3$ & $1.3$ & $1.2$ & $\times 10^{-6}$\\ 
    $2^{-(+)}$ & $1.6$ & $1.3$  & $0.9$ & $\times 10^{-7}$\\ 
\hline     
    $0^{+(-)}$ & $3.0$ & $2.3$ & $1.6$ & $ \times 10^{-4}$\\ 
    $1^{-(+)}$ & $4.9$ & $3.6$ & $2.5$ & $ \times 10^{-6}$\\ 
    $2^{+(-)}$ & $3.3$ & $2.3$ & $1.5$ & $ \times 10^{-4}$\\ 
 
  \end{tabular} 
     \end{center} 
  \caption{$b.r.(\bar{B}^0 \to \pi^- D_H^+(J^{P(C)}) )$ \label{Dpitab}}   
\end{table} 
 
Whilst these are the decays in which the naive factorisation approximation is most  
likely to be correct they are not ideal for hybrid hunting: the $D$-mesons are not  
eigenstates of $C$ and as such cannot have ``exotic'' quantum numbers, the smoking-gun  
signature for a hybrid. However, the fact that for a given $J^{P(C)}$ the hybrid and conventional 
states have \qq~ coupled to $S=0(1)$ and $S=1(0)$ respectively can lead to 
selection guides. This has been noted already for hadronic processes\cite{CP95,CDK02}. 
We shall see that there are further examples in B-decays.  
 
The Class II decay $\bar{B}^0 \to D^0 (n\bar{n})_H$, by contrast, 
can produce exotic quantum numbered hybrids and is not suppressed by any small CKM  
matrix elements. Unfortunately there  
are problems both with factorisation \cite{Stewart} and with our model formulation when  
applied to this decay. Our model has form-factors obtained from the simple wavefunction  
overlaps $\la H | j_L | B \ra$, and although these are reasonable at small $|\vec{q}|$  
where everything is non-relativistic they are at best a qualitative guide 
as the phase-space rises.  
Such subtleties are discussed in \cite{isgw2} where a better fit to the spectrum of  
semi-leptonic $B$ decays is found using a power law form-factor in contrast to the 
``polynomial times exponential" form that we have used. To reduce the uncertainty we calculate 
the ratio of hybrid production and conventional (known) rates, for which much of the 
form factor dependence cancels out. Thus for example, if we propose that there is a $1^{-+}$ hybrid at  
1600 MeV we predict that it will have a branching fraction a little smaller than that of  
$\bar{B}^0 \to D^0 (\rho, \omega)$ i.e. ${\cal O}(10^{-5})$. $(0,2)^{+-}$ exotics have  
potentially even larger rates provided that they have masses somewhat below 3 GeV.  
 
In addition to $B$-meson decays, this formalism can equally well be applied  
to $D$ and $D_s$ decays, though the  
factorised model may not be as accurate here. There is some possibility of  
producing exotic hybrids in the Cabibbo suppressed decay  
$D^0 \to \pi^+ (n\bar{n})_H^-$ if the hybrid is light enough, similarly in the  
channel $D_s \to \pi^+ (s\bar{s})_H$. The numerical branching ratios for  
these processes are strongly dependent on the hybrid mass as they are so  
close to kinematic threshold. The candidate $1^{-+}$ state at 1600 MeV  
has a branching ratio in this decay of ${\cal O}(10^{-7})$  principally  
because it is produced in a $P$-wave with a very small $|\vec{q}|$. 
 
\subsection{$B$ decays to Hybrid and conventional vector} 
 
The matrix element ${\cal M}(B \to V H) = m_V f_V \epsilon_\mu^* \la H 
|(V-A)^\mu|B\ra$, where $\epsilon_\mu$ is the vector meson 
polarisation. We thus have a set of amplitudes depending upon the 
helicity of the vector. Examples of this type of decay include 
$\bar{B}^0 \to \rho^- D^{(*)+} $ (Class I) and the well tested decay 
$B^+ \to J/\psi K^{(*)+}$ (Class II). 
 
As discussed in Section \ref{model}, the $|{\cal M}|^2$ for the weak 
transition $B \to \psi K_H(1^+)$ is expected to have strength $\sim 
13\%$ relative to its ``conventional" counterpart $B \to \psi K(1^+)$. 
Empirically $B^+ \to \psi K(1^+)(1280)$ is the single largest 
mode in $B^+ \to \psi X$ with $b.r.  = (1.8 \pm 0.5)\times 
10^{-3}$ while $B^+ \to \psi K(1^+)(1400) \leq 0.5 \times 10^{-3}$. 
These rates involve both parity conserving (vector) and violating 
(axial) contributions and their relative strengths depend on the 
mixing between the $^3P_1$ and $^1P_1$ basis states.  These rates 
would lead one to expect an order of magnitude $b.r.$ for $B^+ \to 
\psi K_H(1^+) \geq 10^{-4}$. 
This has been reported by us in ref.\cite{cd03}. Here we illustrate 
the calculation. 
 
\subsubsection*{Spin-singlet Hybrids} 
As previously we select the terms in the hybrid currents (Appendix \ref{hybVA}) independent of $\vec{\sigma}$, for which we find terms  
in both $V^\mu_H$ and $A^\mu_H$. We compute separately the amplitudes  
for longitudinal and transverse $J/\psi$: 
 
\paragraph*{Longitudinal $J/\psi$ and $K_H(1^+)$}. 
 
For a longitudinal $J/\psi$, $\epsilon_{\psi}^{\mu} = (|\vec{q}|,0_\perp,E_{\psi})/m_{\psi}$ and so  
\[  
{\cal M}_L = f_\psi \la K_H | |\vec{q}| V^0_H - E_\psi V^3_H | B \ra  
\]   
since only $A_{H\perp}$ are non-zero for spin-singlets. Using the overlaps in eqn.(\ref{xz}) gives 
\[   
{\cal M}_L = -i f_\psi \delta({\cal P}=+)\delta_{m',0}\sqrt{\frac{2b}{3\pi^3}}\sqrt{4m_Bm_K}    
\times   
\]   
\begin{equation}   
\left\{   
 \frac{6|\vec{q}|}{m_d}\left(1 + \frac{E_\psi}{2 m_s}\right) \la j_1 \ra+ \frac{\pi}{2 m_b}\left(\frac{|\vec{q}|^2}{2m_s}   
 +E_\psi\left(\frac{m_b}{m_s}+1\right)\right) \left(\la j_0\ra + \la j_2 \ra \right)    \right\}.   
\label{psil}   
\end{equation} 
Thus only the positive parity state is produced; there is no longitudinally polarised 
$K_H(1^{-(-)})$. 
 
\paragraph*{Transverse $J/\psi$ and $K_H(1^{\pm(\pm)})$}. 
 
For transverse polarisation $\epsilon_{\psi}^{*\mu} = 
(0,\vec{\epsilon}^*)$ and $\vec{\epsilon}^* = \mp (1,\mp 
i,0)/\sqrt{2}$. Both vector and axial can now contribute. The vector 
current leads to a matrix element 
\begin{equation}   
{\cal M}_{\pm}^V = i f_\psi m_\psi \sqrt{4m_Bm_K} \sqrt{\frac{2b}{3\pi^3}}   
 \frac{\pi}{2 m_b} \left(\frac{m_b}{m_s}+1\right) \delta_{m',\pm1} \left\{\delta{({\cal P}=+})   
 \left(\la j_0 \ra - \frac{1}{2}\la j_2 \ra\right) \pm \delta{({\cal P}=-})  \frac{3i}{2} \la j_1 \ra\right\},   
\label{psitv}   
\end{equation}   
and the axial current to 
\begin{equation}  
{\cal M}_{\pm}^A = \pm i f_\psi m_\psi \sqrt{4m_Bm_K} \sqrt{\frac{2b}{3\pi^3}} \frac{\pi}{2m_b} \frac{|\vec{q}|}{2m_s}  \delta_{m',\pm 1}  
\left\{ \delta({\cal P}=+) \left(\la j_0\ra - \frac{1}{2}\la j_2\ra \right)   
 \pm \frac{3i}{2} \delta ({\cal P}=-) \la j_1 \ra  \right\}   
\end{equation} 
Hence the $V_{\mu}$ and $A_{\mu}$ transitions combine to give the 
total matrix element for transition to $1^{+(+)}$ 
\begin{equation}   
{\cal M}_\pm^{V-A}(1^{+(+)}) = i f_\psi m_\psi \sqrt{4m_Bm_K} \sqrt{\frac{2b}{3\pi^3}} \frac{\pi}{2 m_b}   
\left\{ \la j_0\ra - \frac{1}{2}\la j_2\ra \right\} \delta_{m',\pm 1}\left[ \left(1 + \frac{m_b}{m_s}\right)_{(V)} \mp \left(\frac{|\vec{q}|}{2 m_s}\right)_{(A)}  \right]   
\label{kh1+}   
\end{equation} 
The transition to $1^{-(-)}$ has the same structure apart from the 
partial wave contributing as $j_1$, and reads 
\begin{equation} 
 {\cal M}_\pm^{V-A}(1^{-(-)}) = - i f_\psi m_\psi \sqrt{4m_Bm_K} 
\sqrt{\frac{2b}{3\pi^3}} \frac{\pi}{2 m_b} \left\{ \frac{3i}{2} \la 
  j_1\ra \right\} \delta_{m',\pm 1}\left[ \left(1 + 
    \frac{m_b}{m_s}\right)_{(V)} \mp \left(\frac{|\vec{q}|}{2 
      m_s}\right)_{(A)} \right] 
\label{kh1-}   
\end{equation} 
We see that for each of these transitions, the vector current 
dominates. Hence we may anticipate that the transition to $K_H(1^+)$ 
will be relatively large because the dominant $\la V_{\mu} \ra$ 
contributes in $S$-wave; by contrast $K_H(1^-)$ receives its $S$-wave 
from the $|\vec{q}| / m_b$-suppressed $\la A_{\mu} \ra$ while the 
vector current contributes to $P$-waves.  Explicit calculation, below, 
confirms this. 
 
\paragraph*{Longitudinal to transverse ratio}. 
The $1^-$ hybrid is produced in transverse polarisation only. For the 
$1^+$ hybrid both transverse and longitudinal polarisations are 
possible. Compare the longitudinal matrix element, eqn.(\ref{psil})with 
the transverse eqn.(\ref{kh1+}) and remove common factors ($A=f_\psi 
\sqrt{\frac{2 b}{3 \pi^3}} \sqrt{4 m_Bm_K}$) 
\begin{eqnarray*}  
|{\cal M}_L| &=& A \left\{   
 \frac{6|\vec{q}|}{m_d}\left(1 + \frac{E_\psi}{2 m_s}\right) \la j_1 \ra+ \frac{\pi}{2 m_b}\left(\frac{|\vec{q}|^2}{2m_s} + E_\psi\left(\frac{m_b}{ms}+1\right)\right) \left(\la j_0\ra + \la j_2 \ra\right)    \right\} \\  
|{\cal M}_{T(\pm)}| &=& A m_\psi     
\frac{\pi}{2 m_b} \left\{ \la j_0\ra - \frac{1}{2}\la j_2\ra   
   \right\} \left[ \left(1 + \frac{m_b}{m_s}\right)_{(V)} \mp \left(\frac{|\vec{q}|}{2 m_s}\right)_{(A)}  \right].   
\end{eqnarray*}   
Before taking the ratio, consider the actual values of the parameters. 
For a 1.9 GeV Kaon hybrid we have $|\vec{q}|=0.83\mathrm{GeV}$, thus 
$|\vec{q}|/2m_s \approx 0.8$ is negligible next to $(1+m_b/m_s)\approx 
10.5$. The $J/\psi$ is moving non-relativistically such that for this 
$|\vec{q}|$, $E_\psi \approx m_\psi$ is good. Using our variational 
wavefunctions (Appendix \ref{hybham})  we find $\la j_0 \ra \approx 0.58$, $\la j_1 
\ra \approx 0.23$ and $\la j_2 \ra \approx 0.06$, so we can safely 
neglect $\la j_2 \ra$. Thus we find, 
\[  
\left|\frac{{\cal M}_L}{{\cal M}_T}\right| \approx 1 + \frac{12}{\pi} 
\frac{m_b}{m_\psi} \frac{|\vec{q}|}{m_d} \left(\frac{1+ 
    m_\psi/2m_s}{1+m_b/m_s}\right) \frac{\la j_1 \ra}{\la j_0 \ra} 
\approx 3.4, 
\]    
which leads to a longitudinal width fraction, 
\[  
\% \Gamma_L \equiv 100\frac{\Gamma_L}{\Gamma_L +\Gamma_{T(+)} + 
  \Gamma_{T(-)}} \approx 100 \left(1+ \left|\frac{{\cal M}_T}{{\cal 
        M}_L}\right|^2 \right)^{-1} \approx 90 \%. 
\]    
 
This number should not be taken too seriously however. It has proven difficult to accommodate the measured value of the longitudinal width fraction of $B^+ \to J/\psi K^{*+}$ within factorised models and this failure may signal the limitations of this simplistic model. 
  
\subsubsection*{Spin-triplet Hybrids} 
 
Considering the terms in eqns.(\ref{vh0},\ref{vht},\ref{ah0},\ref{aht}) linear in $\vec{\sigma}$ yields matrix elements, 
\begin{eqnarray} 
{\cal M}_L \left(\begin{array}{c} 
                     0^{+(-)}\\ 
                     1^{-(+)}\\ 
                     2^{+(-)} 
                   \end{array} 
               \right) &=&  -i f_\psi \sqrt{4 m_B m_K} \sqrt{\frac{2 b}{\pi^3}} \frac{1}{3} \times  \nonumber \\ 
& & \hspace{-1.7cm}  \left( \frac{6}{m_d} \la j_1 \ra \left(E_\psi + \frac{|\vec{q}|^2}{2 m_s} \right) 
                   \left\{ \begin{array}{c} 
                     -1\\ 
                     0\\ 
                     \sqrt{2} 
                   \end{array} \right\} + 3 |\vec{q}| \frac{\pi}{2m_b} \left( \frac{m_b}{m_s} +1 +\frac{E_\psi}{2 m_s}\right)   \left\{ \begin{array}{c} 
                     - \la j_0 \ra\\ 
                     i \sqrt{\frac{3}{2}} \la j_1 \ra\\ 
                     \frac{1}{\sqrt{2}} \la j_2 \ra 
                   \end{array} \right\}   \right) \\ 
 {\cal M}_L\left(\begin{array}{c} 
                     0^{-(+)}\\ 
                     1^{+(-)}\\ 
                     2^{-(+)} 
                   \end{array} 
               \right) 
         & = &  f_\psi \sqrt{4 m_B m_K} \sqrt{\frac{2 b}{\pi^3}} \times  \nonumber \\ 
& & E_\psi \frac{\pi}{2 m_b} \left(\frac{m_b}{m_s} -1 - \frac{|\vec{q}|^2}{2 m_s E_\psi }\right)   
               \left\{ \begin{array}{c} 
                     \la j_1 \ra\\ 
                     -i \sqrt{\frac{2}{3}} (\la j_0\ra - \frac{1}{2} \la j_2 \ra)\\ 
                     \frac{1}{\sqrt{2}} \la j_1 \ra 
                   \end{array} \right\} \\ 
\hline \nonumber \\ 
{\cal M}_{\pm} \left(\begin{array}{c} 
                     1^{+(-)}\\ 
                     2^{+(-)} 
                   \end{array} 
               \right) 
               &=& i f_\psi \sqrt{4 m_B m_K} \sqrt{\frac{2 b}{\pi^3}} \frac{1}{\sqrt{6}} \times \nonumber \\ 
& & \hspace{- 1.6cm}m_\psi \left( \frac{6}{m_d} \la j_1 \ra \left(1 \mp \frac{|\vec{q}|}{2 m_s} \right)  
                 \left\{ \begin{array}{c} 
                     \pm 1 \\ 
                     1 
                   \end{array} \right\} 
- \frac{\pi}{2 m_b} \left(\frac{m_b}{m_s} -1 \pm \frac{|\vec{q}|}{2 m_s} \right) \left\{ \begin{array}{c} 
                     2 \la j_0 \ra  + \frac{1}{2} \la j_2 \ra\\ 
                     \pm \frac{3}{2} \la j_2 \ra   
                   \end{array} \right\} 
\right) \\ 
{\cal M}_{\pm} \left(\begin{array}{c} 
                     1^{-(+)}\\ 
                     2^{-(+)} 
                   \end{array} 
               \right) 
               &=& f_\psi \sqrt{4 m_B m_K} \sqrt{\frac{2 b}{\pi^3}} \sqrt{\frac{3}{8}} \times \nonumber \\  
& & m_\psi \frac{\pi}{2 m_b} \left(\frac{m_b}{m_s} -1  \pm \frac{|\vec{q}|}{2 m_s} \right) \la j_1 \ra 
                 \left\{ \begin{array}{c} 
                     \pm 1\\ 
                     1 
                   \end{array} \right\} 
\end{eqnarray} 
 
We can use these matrix elements in the width formula 
\[ 
\Gamma(B^+ \to J/\psi K_H^+(J^{P(C)})) = \frac{G_F^2}{16 \pi} \frac{|\vec{q}|}{m_B^2} |V_{bc} V_{cs}|^2 |a_2|^2 \sum_{+,-,L} |{\cal M}|^2 
\] 
to compute branching fractions. 
 
\subsubsection*{Numerical Estimates} 
 
We explicitly evaluate branching fractions and longitudinal rate 
fractions for the exclusive channel $B^+ \to J/\psi K_H^+$ using the 
parameters and wavefunctions given in Appendix \ref{hybham}, the results being 
presented in Table.(\ref{BpsiK}). 
 
\begin{table}[h] 
  \begin{center} 
  \begin{tabular}{c|ccc|c} 
    $m_{\bar{K}_H}$ (GeV) & 1.8 & 2.0 & 2.1 \\ 
    $\substack{\la j_0 \ra \\ \la j_1 \ra \\\la j_2 \ra}$ & $ \substack{0.54 \\ 0.25 \\ 0.08}$ & $ \substack{0.63 \\ 0.20 \\ 0.04}$  & $ \substack{0.69 \\ 0.14 \\ 0.02}$\\ 
\hline 
    $1^{+(+)}$ &$2.9_{92\%}$& $1.2_{79\%}$&$0.6_{60\%}$ & $\times 10^{-4}$\\ 
    $1^{-(-)}$ &$1.3$ &$0.6$ &$0.2$ & $\times 10^{-5}$ \\ 
\hline     
    $0^{-(+)}$ & $5.7$ & $2.8$ & $1.0$ &$ \times 10^{-6}$\\ 
    $1^{-(+)}$ & $6.0_{32\%}$ & $2.5_{19\%}$ & $0.8_{10\%}$ & $\times 10^{-6}$\\ 
    $2^{-(+)}$ & $7.0_{41\%}$ & $3.5_{40\%}$ & $1.3_{40\%}$ & $\times 10^{-6}$\\ 
\hline     
    $0^{+(-)}$ & $9.8$ & $4.1$ & $1.4$ & $ \times 10^{-5}$\\ 
    $1^{+(-)}$ & $3.1_{5\%}$ & $1.7_{11\%}$ & $0.9_{18\%}$ & $ \times 10^{-4}$\\ 
    $2^{+(-)}$ & $2.9_{39\%}$ & $1.1_{39\%}$ & $0.3_{39\%}$ & $ \times 10^{-4}$\\ 
 
  \end{tabular} 
  \end{center} 
  \caption{$b.r.(B^+ \to J/\psi K_H^+(J^{P(C)}) )$. The subscripts are the longitudinal rate fractions.  \label{BpsiK}}   
\end{table} 
 
While fine details of the model may be questioned, the $O(10^{-4})$ 
branching ratio to the hybrids with positive parity appears robust and 
accessible to experiment.  It is intriguing therefore that there is an 
unexplained enhancement at low $q_{\psi}$, corresponding to high mass 
$K$ systems, of this magnitude\cite{psidata}. 
  
While suggestive, it would be premature to claim this as evidence for 
hybrid production.  Radial excitations of the $K(1^+)$ are expected in 
this region, and in the ISGW\cite{isgw} model, extended to exclusive 
hadronic decays and assuming standard factorisation arguments 
\cite{NeuSte}, we find these to have $b.r.  \sim 10^{-4}$, though 
slightly less than the hybrid. 
  
Other conventional strange mesons in this mass range are likely to be 
suppressed due to their high angular momenta, which give powerful 
orthogonality suppressions at small $|\vec{q}|$.  It is the $S$-wave character 
of the hybrid and axial production that drives their significant 
production rates. 
 
The channel $\bar{B}^0 \to D^{0*} (n\bar{n})_H^0$, whilst suffering the 
same theoretical problems as $\bar{B}^0 \to D^{0} n\bar{n}_H^0$ 
considered earlier, is not in principle suppressed by any small 
numbers. CLEO II has observed events $\bar{B}^0 \to D^{0*} \pi^+ \pi^- 
\pi^+ \pi^-$ \cite{edwards} which is a possible end state for 
light-quark hybrids decaying via $a_1^\pm \pi^\mp$ for example. 
 According to the flux-tube breaking model 
of hybrid hadronic decays \cite{CP95} isovector 
$2^{+-},1^{+-},0^{+-},1^{--}$ and isoscalar $1^{-+}, 1^{++}$ hybrids 
have large branching ratios to $a_1 \pi$.

\section{Discussion}  
  
The results of this paper suggest that the flux tube can be excited 
without undue penalty. The physics assumes that a flux tube is indeed 
formed and if subsequent lattice studies should confirm that the 1 fm 
length scale of confinement is intermediate between that of 
perturbative or bag-like gluonic fields and a fully formed flux-tube, 
then some of our results may need to be reassessed and more mature 
modelling developed.  However, this approach seems likely to exhibit 
features that will survive, at least in part, in a more mature 
description. 
  
There is an obvious question as to the reliability of our 
non-relativistic treatment for light flavours. We suggest however that 
the physics of the excitation is probably more general than the 
specific modelling herein. For very massive quarks, where the 
non-relativistic treatment may be justified, the c.m. of the system 
tends to lie on interquark axis $\vec{r} \equiv \vec{r}_Q - 
\vec{r}_{\bar{Q}} $.  For $|\vec{q}| < m_Q$ the recoil of the quark is small 
and the c.m. tends to remain near this axis. As a consequence, we find 
that excitation of the tube is suppressed. As the quark masses are 
reduced, the c.m. of the system can have increasingly large excursions 
from the $\vec{r}$ axis, and the tube is more easily excited. There 
comes a point where the quarks are light and the non-relativistic 
approximations are suspect; however, the physical picture that the 
flux-tube excitation is easier for this situation than for the heavy 
quark case is a physically reasonable extrapolation. So although the 
actual numbers may be debatable, the probability for excitation of the 
flux-tube seems likely to be at least as big as was calculated for 
massive quarks where the non-relativistic approximations still apply. 
The $J^{PC}$ patterns of the results, and the conclusions about charge 
exchange processes being required for the excitation of hybrids in the 
E1 multipole, also seem to be robust. 
  
Our results suggest a way of assessing the flux-tube excitation in a 
lattice QCD approach.  For an E1 transition with no spin-flip, a $\pi$ 
can be excited to axial mesons. The electrically charged hybrid ($a_{1H}^\pm$) and conventional ($b_1^\pm$) axial 
states thus excited have opposite $G$-parities. Thus the relative 
strength of the matrix elements for electromagnetic E1 transition to 
$G$-parity $\pm 1$ gives a measure of the penalty for exciting the 
flux-tube. 
  
It will be interesting to see if lattice QCD simultaneously can 
describe the magnitude of $\la r^2 \ra_{\pi}$, the E1 
amplitudes for exciting $G=\pm 1$ axial mesons, and test the extent to 
which the electric dipole sum rule relating these is saturated. In our 
approximations we see that it is saturated by the $^1P_1$ 
conventional states and their hybrid axial counterparts. If these results are 
verified, then it may be possible to extrapolate to other multipoles, 
relating the modifications of various static properties by the 
flux-tube degrees of freedom to the excitation of hybrid states of 
various $J^{PG}$. 
  
The hybrid configurations with non-exotic $J^{PC}$ will tend to mix 
into the wavefunctions of the conventional mesons. Unless 
 specific observables (such as $\frac{g_A}{g_V}$) 
have anomalous values, it will be hard to make a convincing case for 
such states. If the enhancement in $B \to \psi X$ can be shown to be 
due to a $K_A$ state, this would be very interesting but would not of 
itself be proof of a hybrid. Further measurements, of polarisation or 
$\frac{g_A}{g_V}$ would be needed to distinguish it from a radial 
excitation of a conventional axial.  The main interest is seeking 
rates for states with exotic $J^{PC}$.  Table.(\ref{supertable}) and eqn.(\ref{bJHrate}) 
predict healthy couplings to $\gamma \rho$, which implies that 
photoexcitation of $0^{+-}, 2^{+-}$ off a $\rho$ is feasible. We note 
also that diffractive production of $2^{+-}$ with photon beams is to 
be expected, hence $2^{+-}$ exotic states should be searched for 
in charge exchange and also in diffractive scattering e.g. at either Jefferson 
Lab or HERA. This is discussed further in \cite{cdpirho}.
  
 In the decays of heavy flavours $B \to D_H X$ hides the exotic quantum numbers in the flavour of 
 the $D_H$. However, the ``spin inversion" between conventional and hybrid vector mesons 
 (the hybrid being a quark spin singlet in contrast to the conventional triplet) 
leads to an interesting selection rule: 
  $B^0 \to \pi^- D^+_H (1^{-(-)}) \equiv 0$. 
  Hence observation of a vector $D^*$ in other processes, that is absent in $B$ decay, could 
  be a signature for $D_H(1^-)$. 
   
  The production of an axial strange hybrid in $B \to \psi K_H$ is predicted to have  
  $b.r. \sim O(10^{-4})$ 
  so long as its mass $\leq 2.1 \mathrm{GeV}$. There is a tantalising unexplained enhancement in the data that may be 
  compatible with this and 
  merits further investigation. This process superficially has an analogue in $B \to D^{*0} n\bar{n}_H$, 
  which opens up the available phase space for $n\bar{n}$ to $\sim 3 \mathrm{GeV}$ enabling light flavoured hybrid states with 
  exotic quantum numbers to be produced. Unlike the previous case, however, there is the  
  possibility that rescattering effects and the failure of factorisation could contaminate 
   our analysis 
  of this process. Modulo this caveat, we have $b.r.(B \to D^{*0} (n\bar{n})_H) \sim {\cal O}(10^{-5})$ for 
  $\pi_1(1600)$, if this is the exotic hybrid with $J^{PC} = 1^{-+}$; $(0,2)^{+-}$ exotics have potentially 
  even larger rates provided they have masses below 3 GeV.
   
The exotic $1^{-+}$ resonance $\pi_1(1600)$ could 
  in principle be looked for in $D$ decays. We find a branching ratio for $D \to \pi \pi_1(1600)$ 
  that is $\sim 10^{-4}$ times that for $D \to \pi \rho$.  
Folding in the predicted\cite{cdpirho,pagepirho} $b.r(\pi_1(1600) \to \pi \rho) \sim 20\%$ 
gives a combined branching ratio to $\pi\pi\rho$ via this exotic of $\sim 10^{-8}$. 
This would be a severe challenge even for high statistics studies that may become  
available at CLEO-c or GSI.  
 
 This exotic should have I=0 partners, if it is indeed a hybrid and not some di-meson effect. 
The (\ss)$_H$ state can be produced in leading order $D_s \to \pi s\bar{s}_H$ if its mass is below 1.9 GeV. 
If the nonet is not ideal then there is the hope of some \ss~ content in each of the I=0 states, one 
hopefully light enough to be accessible. Similar to the above, we find a branching ratio 
that is $\sim 10^{-4}$ times that for $D_s \to \pi \phi$. Here again, one is at best at the limits 
of detection. Another potential source of these exotic hybrids is in the decay $B_s \to J/\psi s{\bar s}_H$ which is the $s$-quark spectator analogue of $B \to J/\psi K_H$. We find a branching fraction for a $1^{-+}$ $s {\bar s}$ hybrid at 1.9 GeV to be $\sim 6 \times 10^{-6}$ which is far lower than current statistics can observe.
 
The general formalism developed here can be applied to any current induced transition. 
Examples include gluon emissions (as in \cc~ cascades), diffractive excitation or $\pi$ emission 
where the pion is treated as an effective $\gamma_5$ current. The latter has a long and successful 
history in describing conventional hadron decays\cite{fkr,godisgur} and can now be applied 
analogously to the production or decays of hybrids involving $\pi$. This is described in  
ref. 
\cite{cdpirho}

\section*{Appendices} 

\appendix   
     
\section{Hybrid transition operator for $V_{\mu}$ and $A_{\mu}$ \label{hybVA}}   
   
We perform the non-relativistic reduction of the vector and axial currents   
allowing for flavour changing. We present the particular case of $B(b\bar{d}) \to D(c\bar{d})$ by quark level $b \to c$. In the following $m_D=m_d+m_c$ and $m_B=m_d + m_b$, which are indeed the meson masses in the extreme non-relativistic limit.  
\subsection*{$V^0$}   
   
For the zeroth (time) component of the vector current we have   
\be  
V^0(p_c,p_b;x) = \bar{c}(x) \gamma^0 b(x) \stackrel{\mathrm{N.R.}}{\longrightarrow} e^{i(p_c-p_b)\cdot x}\left\{1 + \frac{\vec{p}_c  
    \cdot \vec{p}_b}{4m_bm_c} + \frac{i}{4} \vec{\sigma} \cdot \left(  
    \frac{\vec{p}_c}{m_c} \wedge \frac{\vec{p}_b}{m_b}\right)+ \ldots \right\}  
\label{v0}   
\ee   
Considering only the terms linear in $\vec{a}$ obtained from utilising the effect of the momentum operator on flux-tube ground state wavefunctions,  
\be \vec{p}|\chi_0\ra \Bigr|_{\vec{a}-\mathrm{cmpt}} = -i \sqrt{\frac{2b}{\pi}}\beta_1 \; \vec{a} |\chi_0 \ra  
\label{pa}   
\ee   
and expanding the plane wave to leading order in $\vec{q}\cdot  
\vec{a}$ we have the effective transition operator to the first hybrid  
excitation  
\be V_H^0 = e^{-i\vec{q}\cdot \vec{r}\frac{m_d}{m_D}}\sqrt{\frac{2b}{\pi^3}}\beta_1\left\{i  
  \vec{q}\cdot \vec{a} \left(\frac{2r}{m_D} + \frac{\pi}{4m_bm_c}  
  \right) - \frac{\pi}{4m_bm_c} \vec{\sigma} \cdot \vec{q} \wedge  
  \vec{a} \right\}  
\label{vh0}   
\ee  
\vskip 0.2in  
\subsection*{$\vec{V}$}   
   
For the spatial vector current we have   
\be   
\vec{V}(p_c,p_b;x) = \bar{c}(x) \vec{\gamma}\, b(x) \stackrel{\mathrm{N.R.}}{\longrightarrow}  e^{i(p_c-p_b)\cdot x}\left\{ \left( \frac{\vec{p}_b}{2m_b} +   
 \frac{\vec{p}_c}{2 m_c}\right)   
-i\vec{\sigma} \wedge \left( \frac{\vec{p}_b}{2m_b} -   
 \frac{\vec{p}_c}{2m_c}\right) + \ldots \right\}   
\label{vt}   
\ee   
and the relevant transition to first excited hybrid becomes   
\be   
\vec{V}_H = e^{-i\vec{q}\cdot \vec{r}\frac{m_d}{m_D}}\sqrt{\frac{2b}{\pi^3}}\beta_1   
\left\{ -i \frac{\pi}{2m_b}\left(\frac{m_b}{m_c}+1\right)  \vec{a} - i \frac{r}{m_cm_D}(\vec{q}\cdot \vec{a})\;\vec{q}   
+ \frac{\pi}{2m_b}\left(\frac{m_b}{m_c}-1 \right) \vec{\sigma} \wedge \vec{a} + \frac{r}{m_cm_D}(\vec{q}\cdot \vec{a})\; \vec\sigma   
\wedge \vec{q}  \right\}   
\label{vht}   
\ee   
\vskip 0.2in   
\subsection*{$A^0$}   
\be   
A^0(p_c,p_b;x) =\bar{c}(x) \gamma^0 \gamma^5 b(x) \stackrel{\mathrm{N.R.}}{\longrightarrow} e^{i(p_c-p_b)\cdot x}\left\{\vec{\sigma} \cdot   
\left( \frac{\vec{p}_b}{2m_b} +   
 \frac{\vec{p}_c}{2m_c}\right)+ \ldots \right\}   
\label{a0}   
\ee   
and the relevant hybrid transition becomes   
\be   
A_H^0 =  e^{-i\vec{q}\cdot \vec{r}\frac{m_d}{m_D}}\sqrt{\frac{2b}{\pi^3}}\beta_1   
\left\{ -i \frac{\pi}{2m_b}\left(\frac{m_b}{m_c} +1\right) \vec{\sigma}\cdot\vec{a} - i \frac{r}{m_cm_D} (\vec{q}\cdot \vec{a})\; \vec{\sigma}\cdot \vec{a} \right\}   
\label{ah0}   
\ee   
\vskip 0.2in   
\subsection*{$\vec{A}$}   
\be   
\vec{A}(p_c,p_b;x) =\bar{c}(x) \vec{\gamma}\gamma^5 b(x) \stackrel{\mathrm{N.R.}}{\longrightarrow} e^{i(p_c-p_b)\cdot x}\left\{\vec{\sigma}   
\left(1 - \frac{\vec{p}_c \cdot \vec{p}_b}{4m_cm_b} \right) +   
\frac{\vec{p_b}(\vec{\sigma} \cdot \vec{p_c}) +\vec{p_c}(\vec{\sigma} \cdot \vec{p_b})}{4m_cm_b}   
- i \frac{\vec{p}_c \wedge \vec{p}_b}{4m_cm_b}   
\right\}   
\label{at}   
\ee   
and the relevant hybrid transition becomes   
\be   
\vec{A}_H = e^{-i\vec{q}\cdot \vec{r}\frac{m_d}{m_D}}\sqrt{\frac{2b}{\pi^3}}\beta_1   
\left\{i (\vec{q}\cdot \vec{a})\; \vec{\sigma}   
\left(\frac{2r}{m_D} - \frac{\pi}{4m_bm_c}  \right) +    
i\frac{\pi}{4m_bm_c}[(\vec{\sigma} \cdot \vec{a})\vec{q} + (\vec{\sigma}\cdot \vec{q})\vec{a}  ]+   
\frac{\pi}{4m_bm_c}   
 \vec{q} \wedge \vec{a}    \right\}   
\label{aht}   
\ee   

\section{How to calculate transition to hybrid \label{hybtrans}}   
   
 We consider the generic 
 \[   
 {\cal M} \equiv \langle \mathrm{hyb}; \pm, m'| {\cal 
   O}_{\mathrm{ref}} |\mathrm{conv}; l,m \rangle = \int \!d^3\vec{r}\! 
 \int\! d^2\vec{a} \; {\cal H^*}{\cal O}_{\mathrm{ref}} {\cal C} 
\]   
where ${\cal O}_{\mathrm{ref}} \equiv \vec{a} \cdot \vec{x}_i$ and the 
$\pm$ is the hybrid state flux-tube polarisation.  In this specific 
example we shall choose $\vec{x}_- \equiv \hat{x} - i\hat{y}$ and 
calculate the matrix element 
 \[   
 \la \chi_1, \pm| \vec{a}\cdot \vec{x}_{-}|\chi_0 \ra \equiv - \la 
 \chi_1,\pm|(a_1+ia_2){\cal D}^{(1)*}_{-+} - (a_1-ia_2) {\cal 
   D}^{(1)*}_{--} |\chi_0 \ra 
 \]   
 where we have introduced the ${\cal D}$ rotation functions shown explicitly for $j=1$ in Table (\ref{Dtable}).  
 
\begin{table}[h] 
\begin{center} 
\begin{math} 
\nonumber 
\bordermatrix{_{m'} \backslash ^m & +1 & 0 & -1\cr  
+1 & \frac{1}{2} (1+\cos \theta) 
    & - \frac{1}{\sqrt{2}} e^{-i \phi} \sin \theta & \frac{1}{2}e^{-2 i \phi}(1 - 
    \cos \theta) \cr  
0 & \frac{1}{\sqrt{2}} e^{i \phi} \sin \theta & 
    \cos \theta & - \frac{1}{\sqrt{2}} e^{-i \phi} \sin \theta \cr  
-1 & \frac{1}{2} e^{2 i \phi} (1 - \cos \theta) & \frac{1}{\sqrt{2}} 
    e^{i \phi} \sin \theta & \frac{1}{2} (1 + \cos \theta) \cr} 
\end{math} 
\end{center} 
\caption{ $ {\cal{D}} ^{(1)}_{m' m} (\phi, \theta, - \phi)$\label{Dtable}} 
\end{table}

The explicit expression for the matrix element is 
 \[   
 -\int d^2\vec{a} \; \beta_1(a_1 \mp ia_2) \left((a_1+ia_2){\cal 
     D}^{(1)*}_{-+} - (a_1-ia_2){\cal D}^{(1)*}_{--}\right) 
 |\chi_0(a_1)|^2 |\chi_0(a_2)|^2 
 \]   
 where the $\chi_0$ are as in (\ref{newchi1}). 
   
 Rewrite $a_1 \pm ia_2 \equiv a^{\pm}$ and the integral becomes 
   \[   
   -\beta_1\int d^2\vec{a} \; a^{\mp} (a^+{\cal D}^{(1)*}_{-+} - 
   a^-{\cal D}^{(1)*}_{--}) |\chi_0(a_1)|^2 |\chi_0(a_2)|^2. 
 \]   
 Now use $a^+a^- = a_1^2 + a_2^2$ and note that $a^{\pm}a^{\pm}$ 
 vanishes under integration.  This brings us to 
 \[   
 -\beta_1\int d^2\vec{a}\; (a_1^2+a_2^2) (\delta^+{\cal D}^{(1)*}_{-+} 
 - \delta^-{\cal D}^{(1)*}_{--}) |\chi_0(a_1)|^2 |\chi_0(a_2)|^2 
 \]   
 Then using the integral 
 \[   
 \int d^2\vec{a}\; (a_1^2+a_2^2) |\chi_0(a_1)|^2 |\chi_0(a_2)|^2 = 
 \frac{\beta_1^2}{\pi}\int d^2\vec{a}\; 
 (a_1^2+a_2^2)e^{-\beta_1^2(a_1^2+a_2^2)} = \beta_1^{-2} 
 \]  
 we finally obtain the essential angular decompositions as follows: 
\begin{eqnarray}   
 \la \chi_1, \pm| \vec{a}\cdot \vec{x}_{-}|\chi_0 \ra &=& -\frac{1}{\beta_1}\left( \delta^+{\cal D}^{(1)*}_{-+} - \delta^- {\cal D}^{(1)*}_{--} \right) \label{axminus} \\  
\la \chi_1, \pm| \vec{a}\cdot \vec{x}_{+}|\chi_0 \ra &=&  
  +\frac{1}{\beta_1}\left( \delta^+{\cal D}^{(1)*}_{++} - \delta^-{\cal D}^{(1)*}_{+-} \right) \label{axplus} \\  
\la \chi_1, \pm| \vec{a}\cdot \hat{z}|\chi_0 \ra &=&  
  -\frac{1}{\sqrt{2}\beta_1}\left( \delta^+{\cal D}^{(1)*}_{0+} - \delta^-{\cal D}^{(1)*}_{0-} \right) \label{az}   
\end{eqnarray}  
  
To proceed from eq.\ref{1s0me} first expand the exponential in terms 
of partial wave angular states $\equiv \sum_L i^L (2L+1) {\cal 
  D}^{(L)*}_{00} j_L(-|\vec{q}|r\frac{\mu}{m_Q})$, contract together 
the three ${\cal D}$ functions 
  
 \[   
 \left[\int \!d\Omega \; {\cal D}^{(1)}_{m',\pm}{\cal 
     D}^{(L)*}_{00}\left( \delta^+{\cal D}^{(1)*}_{-+} - \delta^-{\cal 
       D}^{(1)*}_{--} \right)\right]^*= 
\]  
 \[   
 \frac{4\pi}{3}\la L0;1 -\!\!1|1m' \ra \left(\delta^+\la L0;1 +\!\!1|1 
   +\!\!1\ra - \delta^-\la L0;1-\!\!1|1 -\!\!1 \ra \right). 
 \]   
 and integrate $\int d\Omega$, which gives for the matrix element 
 \[   
 -\frac{1}{\beta_1}\frac{1}{\sqrt 3} \sum_L i^L (2L+1) {_f\!\!\left\la 
     j_L\right \ra_i} \la L0;1 -\!\!1|1m' \ra \left(\delta^+\la L0;1 
   +\!\!1|1 +\!\!1\ra - \delta^-\la L0;1-\!\!1|1 -\!\!1 \ra \right) 
\]   
and finally 
 \[   
  -\frac{1}{\beta_1}\frac{1}{\sqrt 3} 
 \delta_{m',-1}\left[\left(_f{\la j_0 \ra}_i - \frac{1}{2} {_f \la j_2 
       \ra_i}\right) \left(\delta^+ - \delta^-\right) - 
   i\frac{3}{2}{_f \la j_1 \ra_i} \left(\delta^+ + \delta^-\right) 
 \right] 
 \]   
 where we now only need to calculate the radial expectation values of 
 the spherical Bessel functions.   
   
\section{``Dipole'' form in the adiabatic flux-tube model \label{hybmom}}  
  
In the equal mass case we have for the quark positions, 
\begin{equation} 
\vec{r}_{Q,d} = \pm \frac{1}{2} \vec{r} - \frac{br}{\pi m} \sqrt{\frac{2}{N+1}} \sum_{p, \mathrm{odd}} \frac{1}{p} \vec{a}_p. 
\end{equation} 
Specialising to $p=1$\footnote{we are implicitly making an adiabatic approximation so that the phonon modes do not mix, for corrections to such an approximation see \cite{MerlinPat}} we find the momentum of the quarks at lowest non-trivial order in $\frac{br}{m}$, 
\begin{equation} 
\vec{p}_{Q,d} = \pm \vec{p}_{\vec{r}} - \sqrt{2(N+1)} \vec{p}_{\vec{a}} 
\end{equation} 
where $\vec{p}_{\vec{r}}, \vec{p}_{\vec{a}}$ are the momenta conjugate to $\vec{r}, \vec{a}$. 
 
In the usual non-relativistic quark model the matrix element of the lowest order    
electric dipole operator $\sim \frac{\vec{\epsilon} \cdot \vec{p}}{m}$ can be transformed    
using the commutator $ \vec{p} = i m [H,\vec{r}]$ into an explicit dipole    
form $\sim m |\vec{q}| \vec{\epsilon} \cdot \vec{r}$. This commutation relation    
is valid provided the system Hamiltonian can be written in the form    
$H= \frac{p^2}{2 m} + V(r)$, any other dependence of $\vec{p}$ causing    
deviations from this behaviour. We find that this transformation is justified in the adiabatic approximation to the flux-tube model also:

Minimal coupling of the photon field to a quark at $\vec{r}_Q$ moving    
with $\vec{p}_Q$ leads to a convection current    
operator $\sim \vec{p}_Q \cdot \vec{A}(\vec{r}_Q)$. For    
a plane-wave photon field at lowest order in $\vec{q}.    
\vec{r}_Q$ we get an E1 operator $\sim \frac{\vec{\epsilon}\cdot \vec{p}_Q}{m_Q}$.    
We consider the matrix element of this between a ground state meson and a $p=1$    
hybrid state with one excited phonon. Only the $p=1$ term in the sum in $\vec{p}_Q$    
can excite the flux tube to give a non-zero overlap with the hybrid state,    
the $\vec{p}_{\vec{a}_1}$ acting on the ground-state flux-tube wavefunction    
$\chi_{0_1}(a_1^1) \chi_{0_1}(a_1^2)$ to give a factor $i \frac{b\pi}{N+1} \vec{a}_1$.   
 The full matrix element calculated this way comes out to be proportional to  
\be   
\int r^2 dr R_H R_0.   
\ee   
If instead we use the relation $\vec{p}_Q = i m_Q [H, \vec{r}_Q]$ to   
 transform the operator and expand    
 the commutator to get a factor $(E_{(H)} - E_{(0)}) \approx |\vec{q}|$ (exact    
 in the non-rel limit), we can use the $p=1$ term in the sum in $\vec{r}_Q^\perp$   
  to excite the flux tube. This leads to a matrix element proportional to   
\be   
\frac{|\vec{q}|}{\pi} \int r^2 dr R_H r R_0  
\ee   
which appears to be different to what we had previously. If however we assume    
that the hybrid energy is higher than the conventional meson energy by the    
additional energy in the flux tube, $\la \frac{\pi}{r} \ra$ only, which    
seems reasonable in this adiabatic picture where the quarks do not move, then    
$|\vec{q}| \frac{\la r \ra}{\pi}  = \pi \la \frac{1}{r} \ra \frac{\la r \ra}{\pi} \sim 1$   
\footnote{ $\la H|1/r|1S\ra \la 1S|r|H \ra =    
\sum_n \la H|1/r|n\ra \la n|r|H \ra \delta_{n,1S} \approx 1$ if $\la H|1/r|n \neq    
1S\ra \approx 0$ or $\la H|r|n \neq 1S\ra \approx 0$. In a true HO system (i.e    
with just one $\beta$ for all states and $\delta = 1$ for the hybrid) all the matrix    
elements of $r$ vanish except the lowest and the approximation is exact. The deviation    
from this case is presumed to be not too large.}      
and the two matrix elements approximately agree.   
   
For example in the adiabatic potential model described in Appendix(\ref{hybham}) with $f\to \infty$ we have for the charmonium sector,   
\bea   
M_\psi &=& 3143 \mathrm{MeV} \\   
M_{c\bar{c}H} &=& 4202 \mathrm{MeV} \\   
\Rightarrow |\vec{q}| &\approx& 1059 \mathrm{MeV} \\   
\pi \la \frac{1}{r} \ra &\approx& 1157  \mathrm{MeV}   
\eea   
So we see good agreement. 
   
\section{Hybrid Hamiltonian \label{hybham}} 
 
We follow the formulation of the flux-tube model described in \cite{IsgPat}, \cite{MerlinPat}, \cite{MerThesis}. In the appendix of \cite{IsgPat} the authors show explicitly that the spatial wavefunction of a hybrid state with phonon occupation $\{n_{p+}, n_{p-}\}$ can be written 
\[ 
R_{n L N \Lambda}(r) \sqrt{\frac{2L+1}{4 \pi}} {\cal D}^{(L)*}_{m_L \Lambda}(\phi, \theta, -\phi) 
\] 
where the quantum numbers are; $n$, radial; $(L, m_L)$, angular momentum; $N=\sum_{p} (n_{p+}+n_{p-})$; $\Lambda =\sum_{p} (n_{p+}-n_{p-})$. We are only interested in the lightest hybrids which have one phonon excited in the $p=1$ mode and hence $N=1$ and $\Lambda=\pm 1$. In the adiabatic approximation we get a radial Hamiltonian (acting on $r R(r)$) for these hybrids, 
\begin{equation} 
\label{H1} 
H^1 = \frac{1}{2 \mu} \frac{\partial^2}{\partial r^2} + \frac{L(L+1) - \Lambda^2}{2 \mu r^2} + b r + \frac{\pi}{r} (1-e^{-f \sqrt{b} r}) - \frac{\kappa}{r} + c, 
\end{equation} 
whereas for conventional mesons (no phonons) we would get, 
\[ 
H^0 = \frac{1}{2 \mu} \frac{\partial^2}{\partial r^2} + \frac{L(L+1)}{2 \mu r^2} + b r - \frac{\kappa}{r} + c. \]   
The modified angular momentum barrier in the hybrid case has its origin in the $\Lambda=\pm 1$ carried by the phonon in the tube. $b r$ is the mass energy of the string. $\frac{\pi}{r}$ is the excitation energy of the string in the $p=1$ mode, the additional factor multiplying this is put in by hand and is designed to model the fact that at short distances we do not expect ``stringy'' configurations to dominate in QCD. The remaining potential terms $ - \frac{\kappa}{r} +c$ are introduced by hand, the first of which represents one-gluon-exchange dominance at short distances and the second is required to describe the observed meson spectrum. 
 
The parameters, $m_Q, b, \kappa, c$ are chosen to reproduce approximately the observed conventional meson spectrum (up to spin-dependant splittings). We use the following set of values; 
 
$b=0.18 \mathrm{GeV}^2$, $f=1$, $c=-0.7 \mathrm{GeV}$, $m_{u,d}=0.33 \mathrm{GeV}$, $m_{s}=0.55 \mathrm{GeV}$,$m_{c}=1.77 \mathrm{GeV}$, $m_{b}=5.17 \mathrm{GeV}$. $\kappa$ is allowed to run in a reasonable way so that for light mesons ($n \bar{n}, n \bar{s}, s \bar{s}$) $\kappa=1.07$, for heavy-light mesons ($n\bar{c}, s\bar{c}, n\bar{b}, s\bar{b}$) $\kappa = 0.67$ and for heavy-heavy mesons ($c\bar{c}, c\bar{b}, b\bar{b}$) $\kappa$=0.52. 
 
We solve the Schr\"odinger equation variationally using a Harmonic Oscillator basis  
\[ 
R^{HO}_{n,L'}(r)= \sqrt{\frac{2 \Gamma(n)}{\Gamma(n+L'+1/2)}} \beta^{L'+3/2} r^{L'} {\cal L}^{L'+1/2}_{n-1}(\beta^2 r^2) \exp{- \beta^2 r^2 /2}. 
\] 
For conventional states $L'$ here is just the angular momentum quantum number $L$. For the hybrid Hamiltonian the modified angular momentum barrier is cancelled if $L'$ is chosen so it satisfies $L'(L'+1) = L(L+1)-\Lambda^2$, for $L=1, \Lambda=\pm 1$ this means $L' \equiv \delta \approx 0.62$.  
 
\begin{table}[h] 
  \begin{center} 
    \begin{tabular}{l c c  c c  c c} 
        &\multicolumn{2}{c}{$L=0$} & \multicolumn{2}{c}{$L=1$} & \multicolumn{2}{c}{Hybrid}\\ 
        &\multicolumn{2}{c}{$ \frac{2}{\pi^{1/4}} \beta_{1S}^{3/2} e^{-\beta_{1S}^2 r^2 /2}$}  
& \multicolumn{2}{c}{$ \frac{2}{\pi^{1/4}} \sqrt{\frac{2}{3}} \beta_{1P}^{5/2}\; r \; e^{-\beta_{1P}^2 r^2 /2}$}  
& \multicolumn{2}{c}{$ \sqrt{\frac{2}{\Gamma(3/2 + \delta)}} \beta_{H}^{\frac{3}{2} + \delta} r^\delta e^{-\beta_{H}^2 r^2 /2}$}\\ 
        &$\beta_{1S}$ & $M_{1S}$ &$\beta_{1P}$ & $M_{1P}$ &$\beta_{H}$ & $M_{H}$\\ 
\hline 
 $n \bar{n}$ & 0.334 & 0.672 & 0.280 & 1.296 & 0.260 & 1.825 \\ 
 $n \bar{s}$ & 0.370 & 0.780 & 0.306 & 1.386 & 0.275 & 1.961 \\ 
 $s \bar{s}$ & 0.426 & 0.857 & 0.342 & 1.448 & 0.311 & 2.024 \\ 
 $n \bar{c}$ & 0.389 & 2.036 & 0.331 & 2.544 & 0.291 & 3.128 \\ 
 $s \bar{c}$ & 0.469 & 2.092 & 0.387 & 2.571 & 0.335 & 3.195 \\ 
 $c \bar{c}$ & 0.639 & 3.129 & 0.509 & 3.539 & 0.426 & 4.235 \\ 
 $n \bar{b}$ & 0.408 & 5.392 & -     & -     & 0.302 & 6.449 \\ 
 $s \bar{b}$ & 0.510 & 5.455 & -     & -     & 0.355 & 6.563 \\ 
 $c \bar{b}$ & 0.798 & 6.418 & -     & -     & 0.486 & 7.575 \\ 
 $b \bar{b}$ & 1.239 & 9.522 & -     & -     & -     & -     \\ 
      
    \end{tabular} 
    \caption{$\beta$ values/ state masses in GeV} 
    \label{beta} 
  \end{center} 
\end{table}

The Hamiltonians $H^0, H^1$ are found to be diagonal in this basis to a very good approximation if the $\beta$ values listed in Table(\ref{beta}) are used. 
 
Merlin \cite{MerThesis} and Merlin \& Paton \cite{MerlinPat} consider non-adiabatic corrections to this Hamiltonian, the values quoted in Table(\ref{beta}) are actually obtained using this modified Hamiltonian, although the differences in $\beta$ with respect to using $H^1$ are usually small.

\bc   
{\bf Acknowledgements}   
\ec   
   
This work is supported, in part, by grants from the Particle Physics and   
Astronomy Research Council,  and the   
EU-TMR program ``Euridice'', HPRN-CT-2002-00311. We thank P.J.S. Watson for discussions on the dynamics of flux-tubes.

\end{document}